\documentclass[twocolumn,showpacs,showkeys,amsmath,amssymb,superscriptaddress,nofootinbib,floatfix]{revtex4-1}

\usepackage{graphicx}
\usepackage{bm}

\def\vec#1{\mathchoice{\mbox{\boldmath$\displaystyle#1$}}
{\mbox{\boldmath$\textstyle#1$}}
{\mbox{\boldmath$\scriptstyle#1$}}
{\mbox{\boldmath$\scriptscriptstyle#1$}}}


\begin{document}

\title{Wounded nucleon model with realistic nucleon-nucleon collision profile and observables in relativistic heavy-ion collisions}

\author{Maciej Rybczy\'nski}
\email{Maciej.Rybczynski@ujk.edu.pl} 
\affiliation{Institute of Physics, Jan Kochanowski University, PL-25406~Kielce, Poland} 

\author{Wojciech Broniowski} 
\email{Wojciech.Broniowski@ifj.edu.pl}
\affiliation{The H. Niewodnicza\'nski Institute of Nuclear Physics,
  Polish Academy of Sciences, PL-31342 Krak\'ow, Poland}
\affiliation{Institute of Physics, Jan Kochanowski University, PL-25406~Kielce, Poland}

\date{30 August 2011}

\begin{abstract}
We investigate the influence of the nucleon-nucleon collision profile 
(probability of interaction as a function of the nucleon-nucleon impact parameter) in the wounded nucleon model and its extensions 
on several observables measured in relativistic heavy-ion collisions. We find that the participant eccentricity coefficient, $\epsilon^\ast$, 
as well as the higher harmonic coefficients, $\epsilon_n^\ast$, are reduced by 10-20\% for mid-peripheral collisions when the 
realistic (Gaussian) profile is used, as compared to the 
case with the commonly-used hard-sphere profile. Similarly, the multiplicity fluctuations, treated as the function of the number of wounded nucleons in 
one of the colliding nuclei, are reduced by 10-20\%. This demonstrates that the Glauber Monte Carlo codes should necessarily use the realistic 
nucleon-nucleon collision profile in precision studies of these observables. 
Our study is relevant for hydrodynamic calculations which use the Glauber-model initial conditions. 
The Gaussian collision profile is built-in in {\tt GLISSANDO}.
\end{abstract}



\maketitle

\section{Introduction \label{sec:intro}}

The wounded nucleon model \cite{Bialas:1976ed} and its extensions \cite{Kharzeev:2000ph,Broniowski:2007nz} have 
become the basic tool in the analysis of relativistic heavy-ion collisions. The simple idea, based on the 
Glauber approach \cite{Glauber:1959aa,Czyz:1969jg} and finding its foundation in the 
Landau-Pomeranchuk effect \cite{Bialas:2008zza}, leads to proper description of the bulk properties of the data. 
In particular, the centrality dependence of the multiplicities from the RHIC energies ($\sqrt{s_{NN}}=42$~GeV) \cite{Back:2004dy} 
to LHC ($\sqrt{s_{NN}}=2.76$~TeV) \cite{Bozek:2011wa} is properly reproduced in the mixed model \cite{Kharzeev:2000ph}, appending the wounded nucleon model with some particle production from binary collisions.  
Moreover, in the collider experiments it is customary to determine the number of participants as a function of centrality 
with the help of the Glauber Monte Carlo simulations \cite{Wang:1991hta,Werner:1988jr,Broniowski:2007nz,Alver:2008aq,Miller:2007ri}. 
 
The Glauber Monte Carlo calculations have also been used to determine the initial eccentricity of the system. This is a very important quantity, as
it gives rise, via subsequent dynamical evolution of the system, to the observed elliptic flow coefficient, $v_2$. Accuracy of 10\% or 
better is requested in this kind of studies to probe with sufficient accuracy the 
sensitivity to various physical effects, in particular
to the equation of state 
\cite{Chojnacki:2007jc,Broniowski:2008vp,Huovinen:2009yb,Schenke:2010nt}
or the viscosity parameters 
\cite{Muronga:2001zk,Teaney:2003kp,Baier:2006gy,Romatschke:2007mq,Chaudhuri:2006jd,Song:2007fn,Bozek:2007qt,Bozek:2009dw,Schenke:2010rr,Schenke:2011tv} 
of the hydrodynamic medium.  

The statistical nature of the Glauber Monte Carlo unavoidably leads to fluctuations in the distribution of the wounded nucleons, 
which yield the event-by-event eccentricity fluctuations related to the fluctuations of the elliptic
flow~\cite{Aguiar:2000hw,Miller:2003kd,Bhalerao:2005mm,Manly:2005zy,Andrade:2006yh,Voloshin:2006gz,Alver:2006pn,Drescher:2006ca,%
Alver:2006wh,Alver:2006zz,Sorensen:2006nw,Broniowski:2007ft,Alver:2007rm,Hama:2007dq,Voloshin:2007pc,Andrade:2008fa,Hama:2009pk}, the multiplicity
fluctuations as the function of the number of wounded nucleons in 
one of the colliding nuclei (the NA49 experiment setup~\cite{Alt:2006jr}),
and the overall size fluctuations \cite{Broniowski:2009fm}, which can explain the magnitude and 
centrality dependence of the measured transverse-momentum
fluctuations~\cite{Gazdzicki:1992ri,Stodolsky:1995ds,Shuryak:1997yj,Mrowczynski:1997kz,%
Voloshin:1999yf,Baym:1999up,Appelshauser:1999ft,Voloshin:2001ei,Prindle:2006zz,Mrowczynski:2009wk,%
Adams:2003uw,Adamova:2003pz,Adler:2003xq,Adams:2005ka,Grebieszkow:2007xz,na49:2008vb}. 

The concept of wounded nucleons can be applied at non-zero rapidity, where it has been beautifully tested for the deuteron-nucleus 
collisions \cite{Bialas:2004su}, with the conclusion that the rapidity profile of the particle emission from the wounded 
nucleons is asymmetric, peaked in the forward hemisphere, but with substantial emission to the backward direction as well.
Extension of this approach to the nucleus-nucleus case \cite{Gazdzicki:2005rr,Bzdak:2009dr} leads to 
a proper description of the forward-backward multiplicity fluctuations at RHIC 
\cite{Bzdak:2009dr,Bzdak:2009xq,Bzdak:2009zz,Bialas:2010zb} and to a natural 
explanation of the sign and magnitude of the directed flow coefficient, $v_1$ \cite{Bozek:2010bi}. Furthermore, 
theoretical studies of the correlations of the forward and backward elliptic flow with 
fluctuations and the asymmetric emission profile lead to the {\em torque} effect \cite{Bozek:2010vz}, where the 
angle between the forward and backward principal axes of the particle distributions in the transverse plane fluctuates 
substantially on event-by-event basis.
All the above-mentioned features show that the Monte Carlo implementations of the wounded-nucleon approach 
result in rich predictions and allow for numerous dedicated analyses in heavy-ion physics.  
 
In the wounded nucleon model the basic entity is the {\em nucleon-nucleon collision profile}, $p(b)$, defined as the 
probability of inelastic nucleon-nucleon (NN) collision at the impact parameter $b$. It is normalized to the 
total inelastic NN cross section,
\begin{eqnarray}
\int 2\pi b \,db \,p(b)=\sigma_{\rm inel}. \label{eq:inel}
\end{eqnarray}
For the most basic quantities, such as the average number of wounded nucleons in a nucleus-nucleus collision at a given centrality, 
the shape of the NN collision profile is irrelevant. For that reason, in most Glauber Monte Carlo codes it is assumed 
for simplicity that $p(b)$ is simply given the 
step function, $p(b)=\Theta(R-b)$, with $\pi R^2 =\sigma_{\rm inel}$. We refer to this as to the hard-sphere approximation.

The purpose of this paper is to show that for certain important heavy-ion observables (eccentricity coefficients, multiplicity fluctuations, 
triangular deformation \cite{Alver:2010gr,Alver:2010dn,Petersen:2010cw}), the shape of 
the NN collision profile is very much important. In particular, the 
use of a realistic $p(b)$, as determined from the low-angle $pp$ scattering \cite{Bialas:2006qf}, yields noticeably {\em lower} 
eccentricity parameters and multiplicity fluctuations, with the reduction at the level of 10-20\% for the mid-peripheral collisions.
By providing the initial condition, our study is relevant for ``precision'' hydrodynamic calculations, 
probing the equation of state, viscosity parameters, etc.

\section{Basic elements of the model}

The unitarity condition links the elastic NN scattering amplitude, $t_{\rm el}(b)$, with the inelastic collision profile 
(we neglect, as is customary, the small real part of the NN scattering amplitude)
\begin{eqnarray}
t_{\rm el}(b)=1-\sqrt{1-p(b)}. 
\end{eqnarray}
Through passing to the momentum representation one obtains
\begin{eqnarray}
T_{\rm el}(q)=\int d^2 b e^{i \vec{b} \cdot \vec{q}} t_{\rm el}(b).
\end{eqnarray}
Further, from the optical theorem, the total NN cross section is 
\begin{eqnarray}
\sigma_{\rm tot}=2 T_{\rm el}(0), \label{eq:tot}
\end{eqnarray}
the total elastic cross section is the difference 
\begin{eqnarray}
\sigma_{\rm el}=\sigma_{\rm tot}-\sigma_{\rm inel} = \int d^2 b |t_{\rm el}(b)|^2, \label{eq:el} 
\end{eqnarray}
and, finally, the elastic differential cross section is given by
\begin{eqnarray}
\frac{d\sigma_{\rm el}(t)}{dt}=\frac{1}{4\pi} |T_{\rm el}(q)|^2, \label{eq:diff}
\end{eqnarray}
with $t=-q^2$.

\begin{figure}[tb]
\includegraphics[width=.45\textwidth]{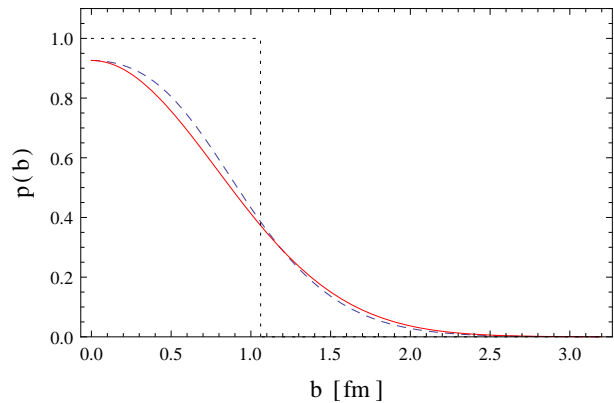}
\caption{(Color online) Comparison of the fit of Ref.~\cite{Bialas:2006qf} (dashed line), formula (\ref{eq:1g}) (solid line),
and the hard-sphere profile (dotted line) for the probability of the inelastic NN collision, $p(b)$, plotted as a function
of the NN impact parameter ($\sigma_{\rm inel}=36$~mb).
\label{fig:profile}} 
\end{figure}

In Ref.~\cite{Bialas:2006qf} the ISR experimental data \cite{Bohm:1974tv,Nagy:1978iw,Amaldi:1979kd,Amos:1985wx,Breakstone:1984te} 
for the total and elastic differential $pp$ cross section
were properly parametrized with a combination of Gaussians (in that quark-diquark picture it is
also possible to reproduce the diffractive minimum in the 
ISR data). Here we use for simplicity a single Gaussian form:
\begin{eqnarray}
p(b)=A e^{-{\pi  A b^2}/{\sigma_{\rm in}^{NN}}}. \label{eq:1g}
\end{eqnarray}
The profiles are compared in Fig.~\ref{fig:profile}. By construction, all fits give the same total inelastic NN cross section.
For the fit of Ref.~\cite{Bialas:2006qf} the elastic NN cross section is 7.8~mb, Eq.~(\ref{eq:1g}) yields 6.8~mb, while
the hard-sphere profile gives 7.4~mb.

The small difference between the fit of Ref.~\cite{Bialas:2006qf} (dashed line) 
and formula~(\ref{eq:1g}) (solid line) is innocuous for the study of the heavy-ion observables carried out in this work.   
The parameter $A$ in Eq.~(\ref{eq:1g}) depends weekly on the collision energy, hence we have set 
\begin{eqnarray}
A=0.92 \label{eq:A}
\end{eqnarray} 
for all the studied collisions.
The hard-sphere profile, shown in Fig.~\ref{fig:profile} with a dotted line, is, of course, much different from the Gaussian 
profile. While the shape of $p(b)$ does not affect certain observables, it leads, as shown in the following Sections, 
to noticeable effects in the description of the shape of the initial fireball or in fluctuations of various quantities.

\section{Results of the Monte-Carlo simulations}

The presented results have been obtained with {\tt GLISSANDO} \cite{Broniowski:2007nz}.
The nuclear distributions are generated according to appropriate Woods-Saxon densities.  
The correlations from the NN repulsion are generated in a standard way by precluding the 
centers of nucleons to be closer to one another than the expulsion distance $d$. We 
use $d=0.9$~fm, which reproduces properly \cite{Broniowski:2010jd} the realistic central correlations
implemented in Refs.~\cite{Alvioli:2009ab,Alvioli:2010yk}.

\begin{figure}[tb]
\includegraphics[width=.5\textwidth]{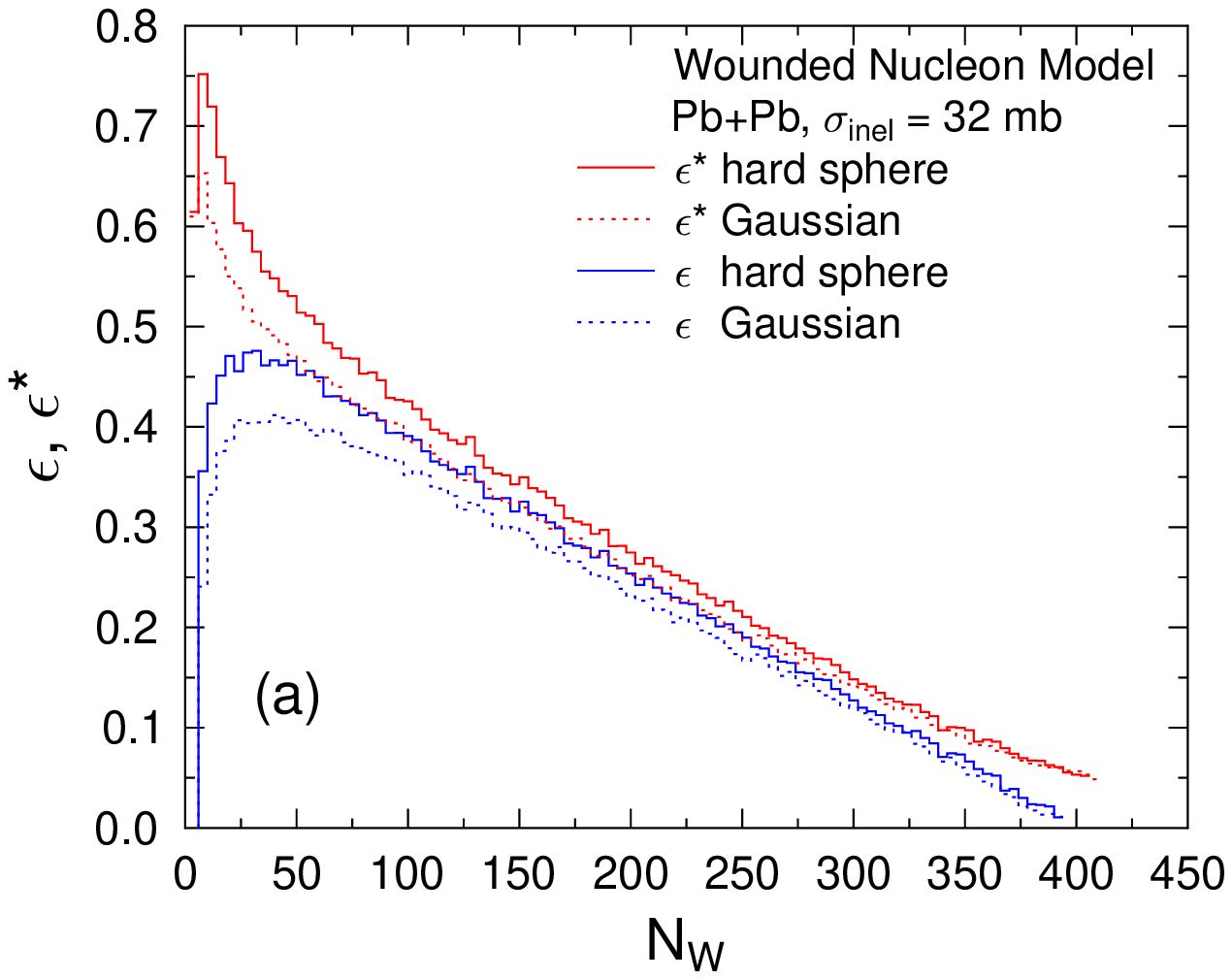}\\
\includegraphics[width=.5\textwidth]{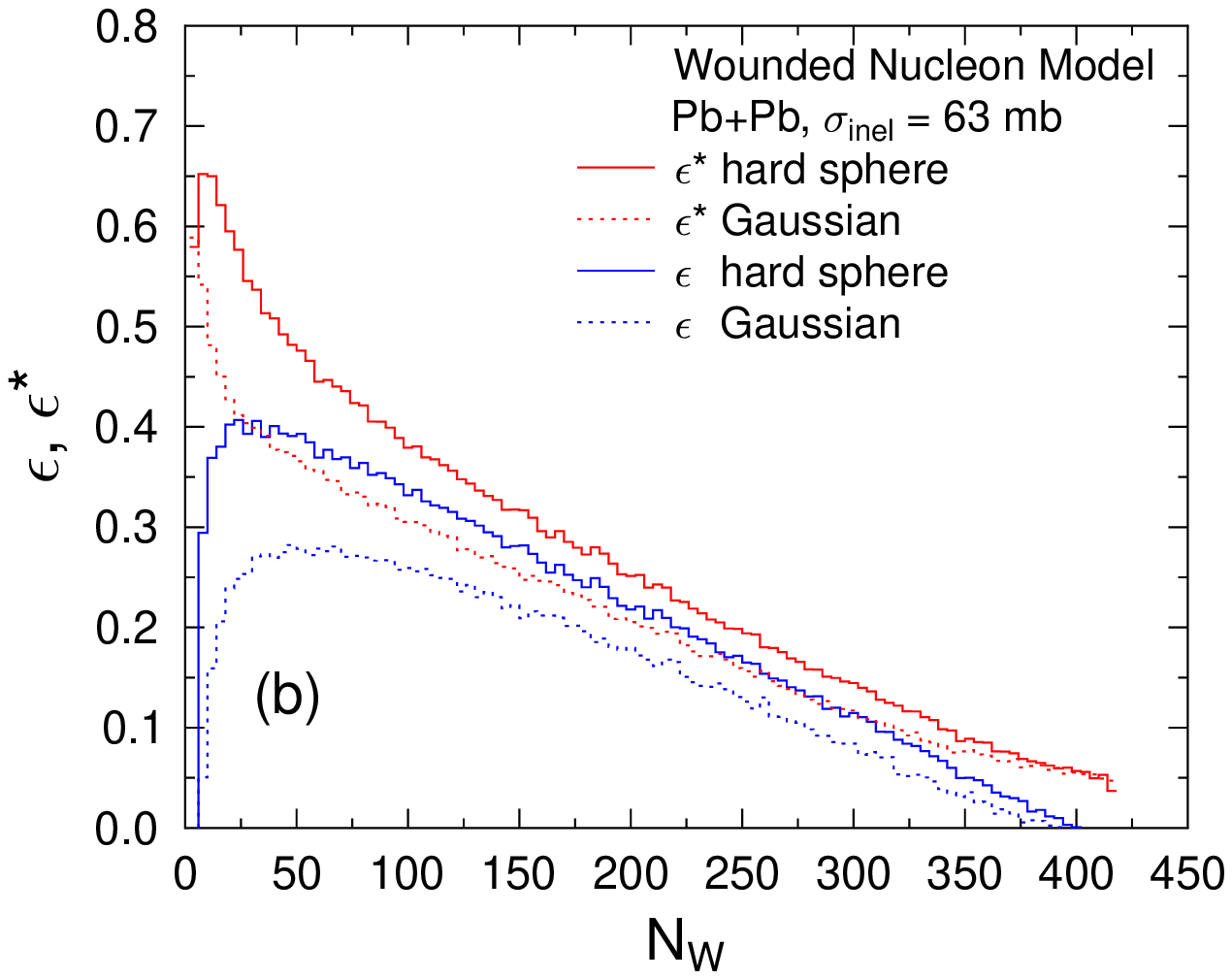}
\caption{(Color online) The event-averaged values of the azimuthal eccentricity $\varepsilon$ (two lower lines) and 
the participant eccentricity $\varepsilon^\ast$ (two upper lines) for the hard-sphere collision profile (solid lines) and the Gaussian collision profile (dashed lines), evaluated in the wounded nucleon model for Pb+Pb collisions and plotted against the total number of wounded nucleons, $N_W$. The panel (a) corresponds to  $\sigma_{\rm inel}=32$~mb (top SPS energy) and the panel (b) to $\sigma_{\rm inel}=63$~mb (LHC). 
\label{fig:eps1}} 
\end{figure}

\subsection{Eccentricity}

The azimuthal eccentricity is defined as the second Fourier moment of the distribution of the wounded nucleons in the transverse plane. 
One defines the eccentricity coefficient, $\varepsilon$, with respect to the axes defined by the reaction plane (the 
coordinate $x$ lies in the reaction plane, and $y$ is perpendicular to it),
\begin{eqnarray}
\varepsilon=\frac{\sigma_y^2 - \sigma_x^2}{\sigma_y^2 + \sigma_x^2}. 
\end{eqnarray}
Since the principal axes of the distribution fluctuate event-by-event, a more appropriate measure is the value of the 
so-called participant eccentricity, evaluated with respect to the principal axes in each event
\cite{Alver:2006zz}. Then
\begin{eqnarray}
\varepsilon^\ast=\frac{\sqrt{\left ( \sigma_y^2-\sigma_x^2 \right )^2+4\sigma_{xy}^2}}{\sigma_x^2+\sigma_y^2}. \label{epss}
\end{eqnarray}
where $\sigma_x^2$, $\sigma_y^2$ and $\sigma_{xy}$ are the variances and covariance of the nucleon distribution in a given event, respectively.

In the description of the evolution of the system involving hydrodynamics 
\cite{Teaney:2000cw,Muronga:2001zk,Teaney:2003kp,Kolb:2003dz,Hama:2005dz,Huovinen:2006jp,%
Baier:2006gy,Romatschke:2007mq,Chaudhuri:2006jd,Song:2007fn,Bozek:2007qt,Bozek:2009dw,Hirano:2005xf,%
Hirano:2002ds,Broniowski:2008vp,Huovinen:2009yb,Bozek:2009ty,Schenke:2010nt,Schenke:2010rr,Schenke:2011tv}, 
the initial azimuthal eccentricity of the system 
is a very important quantity, as it influences the value of the elliptic flow coefficient, $v_2$ \cite{Ollitrault:1992bk}. 
Moreover, it is believed it can be used to probe the dynamical properties of the system, such as the equation of state 
\cite{Chojnacki:2007jc,Broniowski:2008vp,Huovinen:2009yb,Schenke:2010nt}
or the viscosity parameters 
\cite{Muronga:2001zk,Teaney:2003kp,Baier:2006gy,Romatschke:2007mq,Chaudhuri:2006jd,Song:2007fn,Bozek:2007qt,Bozek:2009dw,Schenke:2010rr,Schenke:2011tv}. 
Thus the initial azimuthal eccentricity, supplied as the initial condition 
to the hydro codes, is an important ingredient. 
The fact that $\varepsilon^\ast > \varepsilon$ makes it easier to develop larger $v_2$, thus the hydrodynamic evolution may be 
somewhat shorter. This, it turn, is favorable for the uniform description of the femtoscopic variables \cite{Broniowski:2008vp}.  

From the methodological point of view, in order to study the details of the system dynamics one should start from the 
most realistic initial condition. Certainly, it is not clear what exactly should be used here, as we are still far from explaining 
the earliest-stage dynamics of the system formed in relativistic heavy ion collisions.  Other approaches, such as the Color Glass 
Condensate (CGC) \cite{Iancu:2000hn,Ferreiro:2001qy} provide a 
different initial condition \cite{Hirano:2005xf,Drescher:2006pi} for hydrodynamics than the Glauber models. However, having 
chosen the Glauber approach, we should use the most realistic ingredients. There is no reason to use the simplistic 
hard-sphere wounding profile, since the use of the realistic profile (as inferred form the pp scattering data) is equally simple, 
while the differences certain for observables at the level of 10\%.

In Fig.~\ref{fig:eps1} we show the results of {\tt GLISSANDO} \cite{Broniowski:2007nz} for $\varepsilon$ and $\varepsilon^\ast$, obtained 
in the wounded nucleon model for the Gaussian wounding profile of Eq.~(\ref{eq:1g}) (dashed lines) and the hard-sphere 
profile (solid lines). We note that the use of the realistic (Gaussian) profile rather than the hard-sphere 
profile results in lower values of the eccentricity 
parameters. The reduction for intermediate centralities is at the level of $10\%$ for the SPS case of $\sigma_{\rm inel}=32$~mb and at the level of $20\%$ for the LHC case of $\sigma_{\rm inel}=63$~mb. In fact, the reduction due to the realistic profile brings down $\varepsilon^\ast$ more-less to the value of $\varepsilon$ evaluated with the conventional hard-sphere profile. Thus, the enhancement due to 
the participant geometry is ``lost'' by the use of the proper wounding profile!   
  
\begin{figure}[tb]
\includegraphics[width=.5\textwidth]{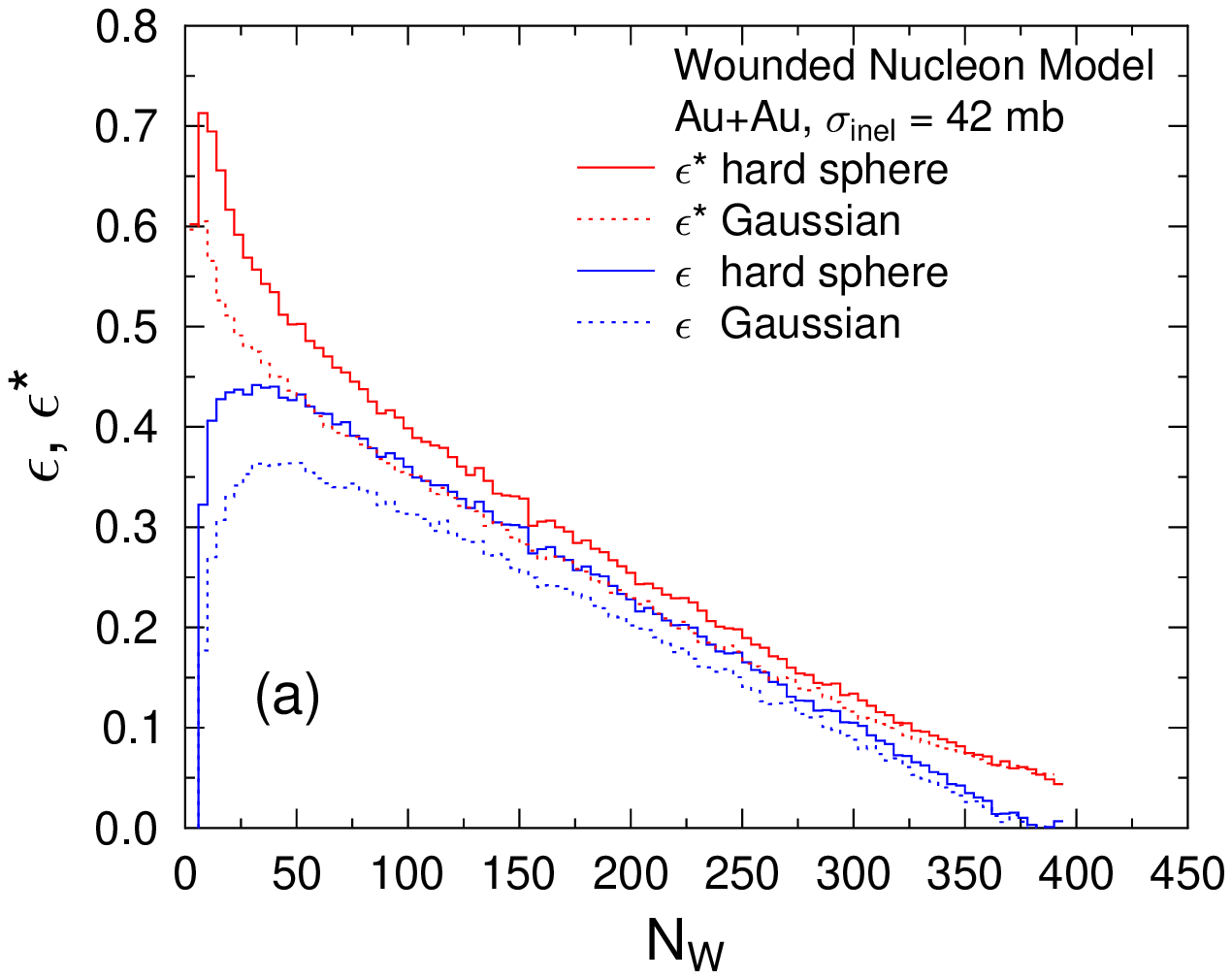}\\
\includegraphics[width=.5\textwidth]{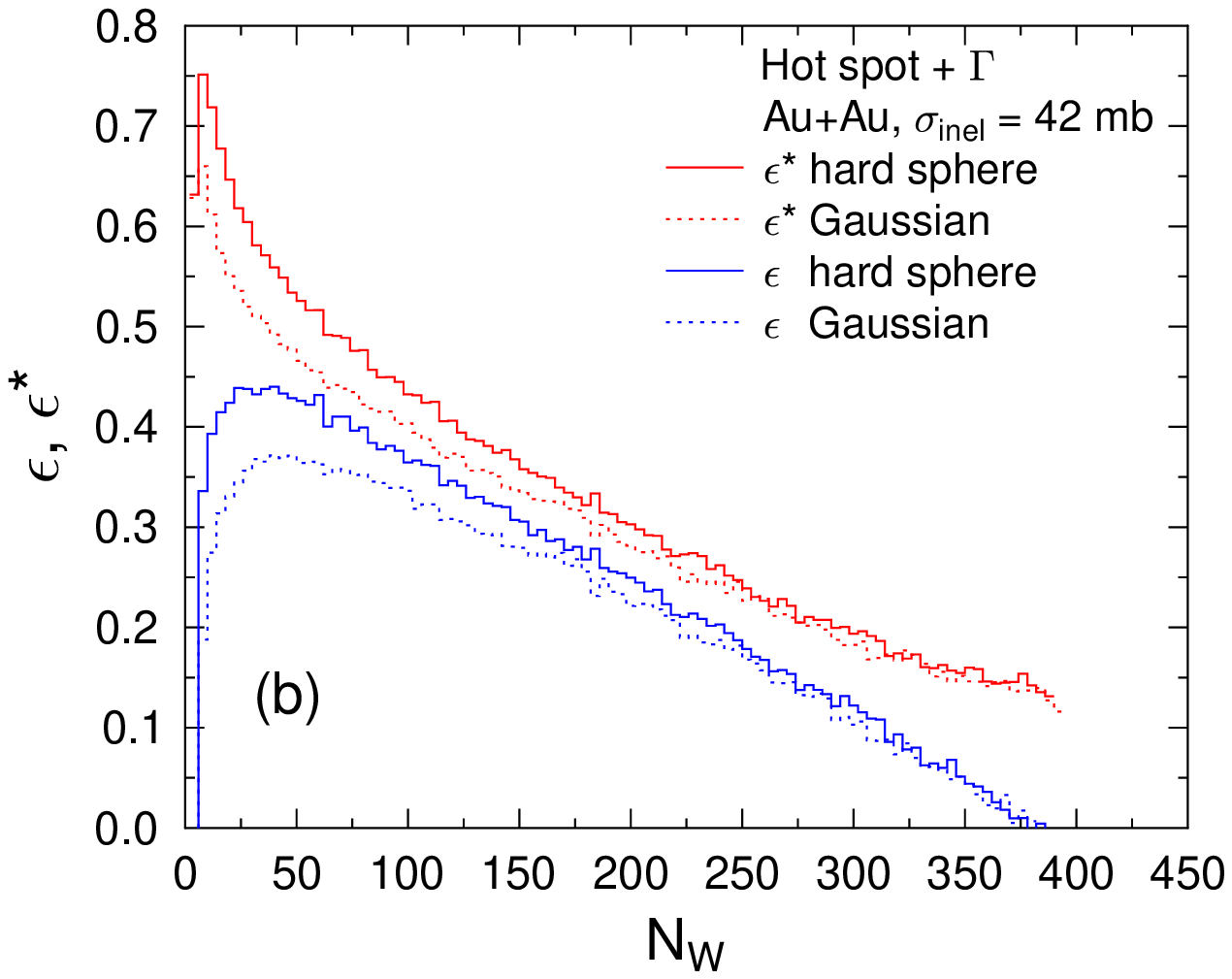}
\caption{(Color online) Same as Fig.~\ref{fig:eps1} for the case of Au+Au collisions with $\sigma_{\rm inel}=42$~mb (top RHIC energy). Panel (a): wounded nucleon model, panel (b): the ``hot spot + $\Gamma$ model''. \label{fig:hsg}} 
\end{figure}

The described quenching effect is easy to understand on geometric grounds. Unlike the hard-sphere 
case, the Gaussian profile has a tail extending to large values of $b$ (cf.~Fig.~\ref{fig:profile}). Therefore there is a certain probability that a nucleon from nucleus $A$ wounds a nucleon from nucleus $B$ which is ``far away'' in the transverse plane. Since there are more nucleons in the $x$-direction (parallel to the reaction plane) than in the $y$-direction (perpendicular to the reaction plane), with the Gaussian profile more nucleons in the $x$ direction are wounded, hence the reduction effect.

\begin{figure}
\includegraphics[width=.5\textwidth]{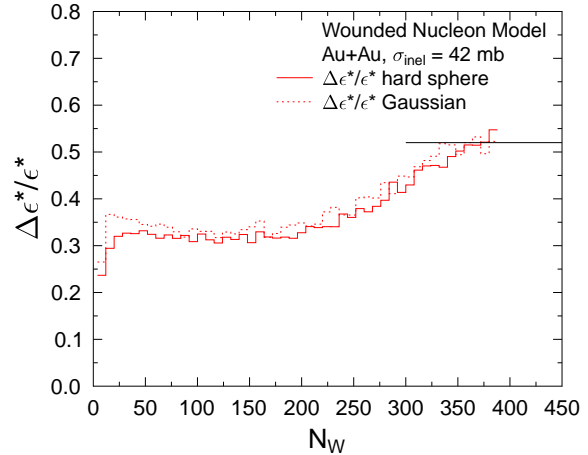}
\caption{(Color online) The event-by-event scaled standard deviation $\Delta \varepsilon^\ast/\varepsilon^\ast$ for the Au+Au collisions with $\sigma_{\rm inel}=42$~mb (top RHIC energy), evaluated for the hard-sphere profile (solid line) and the Gaussian profile (dashed line), plotted as a function of the total number of wounded nucleons, $N_W$. The horizontal line shows the limit of $\sqrt{4/\pi-1}$ derived in~\cite{Broniowski:2007ft}.
\label{fig:fleps}} 
\end{figure}

Next, we check how the effect depends on the variant of the Glauber-like model of the early phase of the reaction. The wounded nucleon model is the simplest model of this kind. Extensions involve admixing particles from the binary collisions (the so-called mixed model of Ref.~\cite{Kharzeev:2000ph,Back:2001xy,Back:2004dy}) or overlaying a distribution of particles over the sources (here the ``source'' means a wounded nucleon or a binary collision). 
The model which leads to large fluctuations is the hot-spot model \cite{Gyulassy:1996br,Broniowski:2007nz}. It assumes that the cross section for a semi-hard binary collision producing a hot spot is small, $\sigma_{\rm bin} \simeq 2$~mb, but when such a rare collision occurs it produces a large amount of transverse energy equal to \mbox{$\alpha\sigma_{\rm w}/\sigma_{\rm bin}$}. Then, we overlay the $\Gamma$ distribution on the produced sources, as described in Ref.~\cite{Broniowski:2007nz}.
The purpose of using this model is, as mentioned, to investigate a case with large statistical fluctuations.

The results are shown in Fig.~\ref{fig:hsg}, where we show the effect of the reduction of the azimuthal eccentricity 
for the case of the highest RHIC energy, comparing the wounded nucleon model (top) and the hot-spot+$\Gamma$ model (bottom). We note that for both cases the reduction is similar, hence the effect does not depend on the particular Glauber-like model used for the analysis. 

In Fig.~\ref{fig:fleps} we show a measure of the event-by-event fluctuations of $\varepsilon^\ast$, namely, the scaled standard deviation of this quantity, investigated in 
Refs.~\cite{Aguiar:2000hw,Miller:2003kd,Bhalerao:2005mm,Manly:2005zy,Andrade:2006yh,Voloshin:2006gz,Alver:2006pn,%
Alver:2006wh,Sorensen:2006nw,Broniowski:2007ft,Alver:2007rm,Hama:2007dq,Voloshin:2007pc} in connection with the fluctuations of the elliptic 
flow. We note a small increase of the fluctuations for this quantity when the Gaussian collision profile is used compared to the hard-sphere case.

\subsection{Higher azimuthal Fourier components}

\begin{figure}
\includegraphics[width=.5\textwidth]{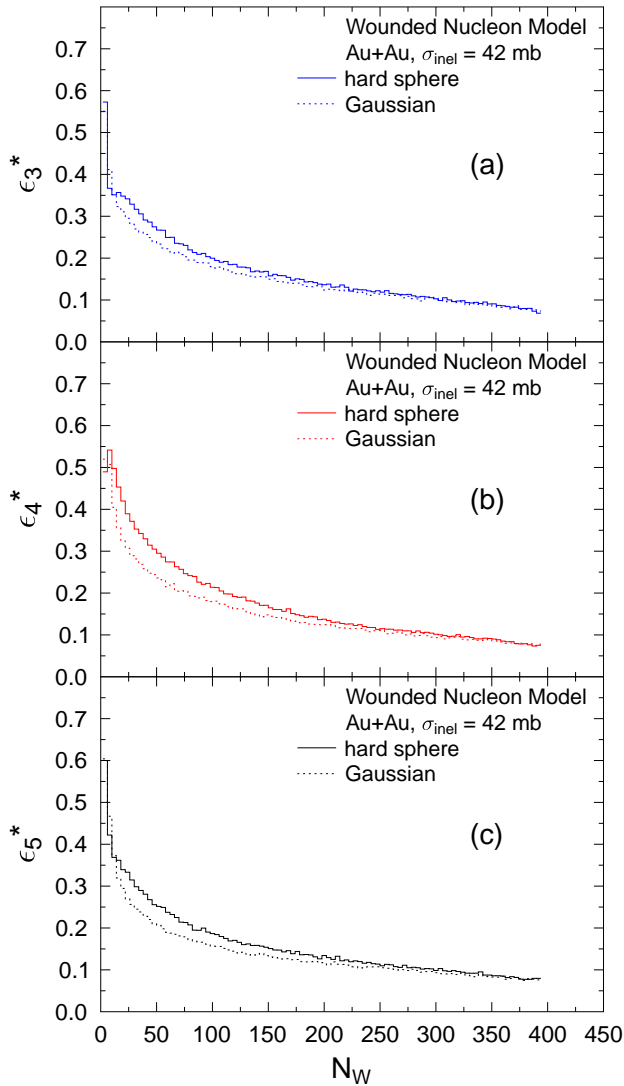}
\caption{(Color online) The wounded-nucleon model predictions for the third ( panel (a)), fourth (panel (b)), and fifth (panel (c)) Fourier 
coefficients of the azimuthal distribution. See the text for details. \label{fig:higher}} 
\end{figure}

A similar reduction effect occurs for the higher Fourier components of the azimuthal distribution of sources. 
The coefficients of rank $n$,  $\varepsilon_n^\ast$,  are defined in each event as \cite{Broniowski:2007ft,Alver:2010gr}
\begin{eqnarray}
\varepsilon_n^\ast = \frac{\sum_i \rho_i^2 \cos[n(\phi_i-\phi^\ast)]}{\sum_i \rho_i^2},  
\end{eqnarray}
where $i$ runs over the number of sources (wounded nucleons or binary collisions) in each event, $\rho_i$
is the distance of the source from the center of mass of the fireball,  and $\phi_i$ is its azimuthal angle. 
The angle $\phi^\ast$ is adjusted in each event in such a way as to maximize 
$\varepsilon_n^\ast$, which gives the condition
\begin{eqnarray}
\tan \phi^\ast = \frac{\sum_i \rho_i^2 \sin(n\phi_i)}{\sum_i \rho_i^2 \cos(n\phi_i)}.
\end{eqnarray}
For $n=2$ the above definition overlaps with Eq.~(\ref{epss}). 
The odd rank coefficients arise entirely due to statistical fluctuations \cite{Broniowski:2007ft}.

The hydrodynamic evolution, when started from initial distributions carrying $\varepsilon_n^\ast$, generates the  
collective flow coefficients $v_n$ \cite{Alver:2010dn,Petersen:2010cw}. They are important in the interpretation of various observed phenomena, 
such as the ridge \cite{Takahashi:2009na,Sorensen:2011xw}. 

In Fig.~\ref{fig:higher} we show the wounded-nucleon predictions for $\varepsilon_n^\ast$ with $n=3$, 4, and 5. 
Similarly to the $n=2$ case, we observe a sizable reduction when the Gaussian collision profile is used. For mid-peripheral 
collisions it is at the level of 15\%. Therefore the realistic NN collision profile should be included in the analyses of the triangular and higher-harmonic flow using the Glauber approach.

\subsection{Multiplicity fluctuations}

Another important class of phenomena studied in relativistic heavy-ion collision are the multiplicity fluctuations \cite{Aggarwal:2001aa, Alt:2006jr, Adare:2008ns}, which carry information on the  NN correlations. We have tested several reactions, in particular those to be investigated in the NA61 experiment \cite{Gazdzicki:2008kk,:2009vy,:2009vj,Gazdzicki:2011fx,Gazdzicki:1322135}. 
The basic measure of the multiplicity fluctuations is the {\em scaled variance} of the wounded nucleons in the target, 
\begin{eqnarray}
\omega_{\rm targ} = \frac{\rm var(N_{\rm targ})}{\langle N_{\rm targ} \rangle}, \label{eq:omega}
\end{eqnarray}
plotted as a function of the wounded nucleons of the projectile, which are measured experimentally in the NA49 or NA61 setup.

\begin{figure}
\includegraphics[width=.5\textwidth]{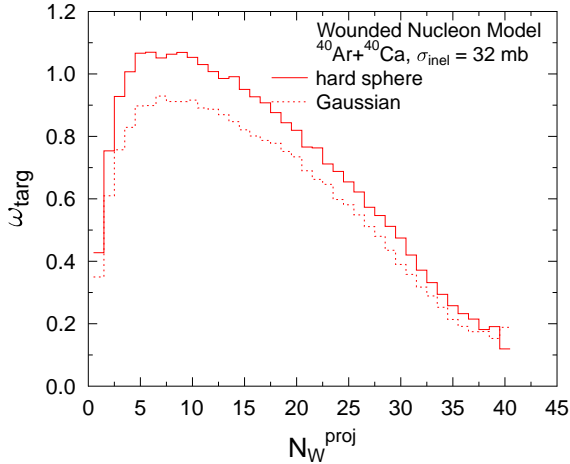}
\caption{(Color online) The multiplicity-fluctuations measure, $\omega_{\rm targ}$, for Ar+Ca collisions with $\sigma_{\rm inel}=32$~mb, evaluated for the hard-sphere collision profile (solid line) and the Gaussian collision profile (dashed line), plotted as a function of wounded nucleons in the projectile, $N_{W}^{\rm proj}$.
\label{fig:arca} }
\end{figure}

A typical result is shown in Fig.~\ref{fig:arca} for the Ar+Ca collision with $\sigma_{\rm inel}=32$~mb. We note 
a reduction of $\omega_{\rm targ}$ when the Gaussian profile is used compared to the hard-sphere case. The effect is at the level 
of 15\%. A similar effect is found for other combinations of the colliding nuclei.

\begin{figure}
\includegraphics[width=.5\textwidth]{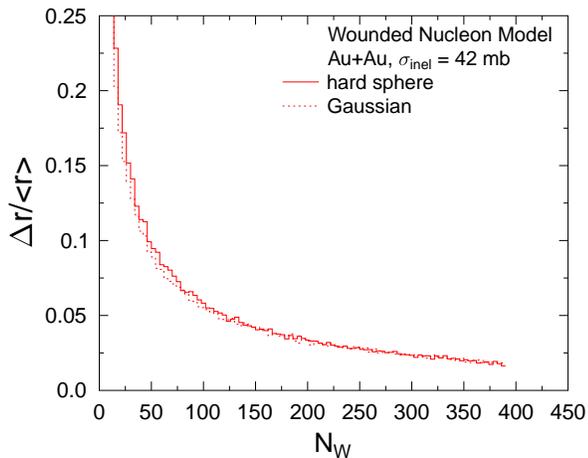}
\caption{(Color online) The event-by-event scaled standard deviation of the size variable $r$ for Au+Au collisions at 
$\sigma_{\rm inel}=42$~mb (top RHIC energy) for the hard-sphere collision profile (solid line) and the Gaussian collision profile (dashed line). \label{fig:r}} 
\end{figure}

\subsection{Other results}

We have also carried out other analyses done in heavy-ion physics, using the 
Gaussian profile in {\tt GLISSANDO}~\cite{Broniowski:2007nz}. The forward-backward multiplicity correlations \cite{:2009dqa,Back:2006id,Bzdak:2009dr,Bialas:2011bz} are 
not affected by the choice of the NN collision profile. The same is true of the size fluctuations 
studied in Ref.~\cite{Broniowski:2009fm}, which influence the transverse-momentum fluctuations. The 
size variable is defined in each event as
\begin{eqnarray}
r=\sum_i \rho_i,
\end{eqnarray}
where $i$ runs over the sources and $\rho_i$ is the distance from the center of mass of the fireball.  
The result for the event-by-event scaled standard deviation of $r$ is shown in Fig.~\ref{fig:r}. We note a very small effect 
from the choice of the NN collision profile, with a small reduction when the Gaussian profile is used.

\section{Conclusions \label{sec:concl}}

We conclude be reiterating our main results: 
\begin{itemize}
\item The Glauber Monte Carlo codes for relativistic heavy-ion collisions 
should definitely use the realistic NN collision profile, as it affects 
important observables in a sizable way. 

\item The eccentricity parameters, $\epsilon^\ast$, and the higher rank parameters, $\epsilon_n^\ast$,
are reduced by 10-20\% when the realistic (Gaussian) NN collision profile is used, compared to the case with the hard-sphere 
profile.

\item The multiplicity fluctuations, as measured in the NA49 or NA61 experimental setup, are reduced by 10-20\% when the   
realistic NN collision profile is used compared to the hard-sphere case.

\item The initial condition following from the realistic wounding profile should be used in hydrodynamic studies. 

\end{itemize}

We note that {\tt GLISSANDO} 
\cite{Broniowski:2007nz}\footnote{\tt http://www.ujk.edu.pl/homepages/mryb/GLISSANDO/index.html} 
has a built-in realistic (Gaussian) NN collision profile and can be readily used for those analyses where 
sensitivity to the NN collision profiles appears.

\bigskip

One if us (WB) wishes to thank Adam Bzdak for useful discussions.

\medskip

Supported by the Polish Ministry of Science and Higher Education, grants N~N202~263438 and N~N202~249235.

%


\begin{thebibliography}{100}%
\makeatletter
\providecommand \@ifxundefined [1]{%
 \ifx #1\undefined \expandafter \@firstoftwo
 \else \expandafter \@secondoftwo
\fi
}%
\providecommand \@ifnum [1]{%
 \ifnum #1\expandafter \@firstoftwo
 \else \expandafter \@secondoftwo
\fi
}%
\providecommand \enquote [1]{``#1''}%
\providecommand \bibnamefont  [1]{#1}%
\providecommand \bibfnamefont [1]{#1}%
\providecommand \citenamefont [1]{#1}%
\providecommand\href[0]{\@sanitize\@href}%
\providecommand\@href[1]{\endgroup\@@startlink{#1}\endgroup\@@href}%
\providecommand\@@href[1]{#1\@@endlink}%
\providecommand \@sanitize [0]{\begingroup\catcode`\&12\catcode`\#12\relax}%
\@ifxundefined \pdfoutput {\@firstoftwo}{%
 \@ifnum{\z@=\pdfoutput}{\@firstoftwo}{\@secondoftwo}%
}{%
 \providecommand\@@startlink[1]{\leavevmode}%
 \providecommand\@@endlink[0]{}%
}{%
 \providecommand\@@startlink[1]{%
  \leavevmode
  \pdfstartlink
   attr{/Border[0 0 1 ]/H/I/C[0 1 1]}%
   user{/Subtype/Link/A<</Type/Action/S/URI/URI(#1)>>}%
  \relax
 }%
 \providecommand\@@endlink[0]{\pdfendlink}%
}%
\providecommand \url  [0]{\begingroup\@sanitize \@url }%
\providecommand \@url [1]{\endgroup\@href {#1}{\urlprefix}}%
\providecommand \urlprefix [0]{URL }%
\providecommand \Eprint[0]{\href }%
\@ifxundefined \urlstyle {%
  \providecommand \doi [1]{doi:\discretionary{}{}{}#1}%
}{%
  \providecommand \doi [0]{doi:\discretionary{}{}{}\begingroup
  \urlstyle{rm}\Url }%
}%
\providecommand \doibase [0]{http://dx.doi.org/}%
\providecommand \Doi[1]{\href{\doibase#1}}%
\providecommand \bibAnnote [3]{%
  \BibitemShut{#1}%
  \begin{quotation}\noindent
    \textsc{Key:}\ #2\\\textsc{Annotation:}\ #3%
  \end{quotation}%
}%
\providecommand \bibAnnoteFile [2]{%
  \IfFileExists{#2}{\bibAnnote {#1} {#2} {\input{#2}}}{}%
}%
\providecommand \typeout [0]{\immediate \write \m@ne }%
\providecommand \selectlanguage [0]{\@gobble}%
\providecommand \bibinfo [0]{\@secondoftwo}%
\providecommand \bibfield [0]{\@secondoftwo}%
\providecommand \translation [1]{[#1]}%
\providecommand \BibitemOpen[0]{}%
\providecommand \bibitemStop [0]{}%
\providecommand \bibitemNoStop [0]{.\EOS\space}%
\providecommand \EOS [0]{\spacefactor3000\relax}%
\providecommand \BibitemShut [1]{\csname bibitem#1\endcsname}%
\bibitem{Bialas:1976ed}%
  \BibitemOpen
  \bibfield{author}{%
  \bibinfo {author} {\bibfnamefont{A.}~\bibnamefont{Bia\l{}as}}, \bibinfo
  {author} {\bibfnamefont{M.}~\bibnamefont{Bleszy\'nski}},\ and\ \bibinfo
  {author} {\bibfnamefont{W.}~\bibnamefont{Czy\.z}},\ }%
  \bibfield{journal}{%
  \bibinfo {journal} {Nucl. Phys.}\ }%
  \textbf{\bibinfo {volume} {B111}},\ \bibinfo {pages} {461} (\bibinfo {year}
  {1976})%
  \bibAnnoteFile{NoStop}{Bialas:1976ed}%
\bibitem{Kharzeev:2000ph}%
  \BibitemOpen
  \bibfield{author}{%
  \bibinfo {author} {\bibfnamefont{D.}~\bibnamefont{Kharzeev}}\ and\ \bibinfo
  {author} {\bibfnamefont{M.}~\bibnamefont{Nardi}},\ }%
  \bibfield{journal}{%
  \bibinfo {journal} {Phys. Lett.}\ }%
  \textbf{\bibinfo {volume} {B507}},\ \bibinfo {pages} {121} (\bibinfo {year}
  {2001})%
  \bibAnnoteFile{NoStop}{Kharzeev:2000ph}%
\bibitem{Broniowski:2007nz}%
  \BibitemOpen
  \bibfield{author}{%
  \bibinfo {author} {\bibfnamefont{W.}~\bibnamefont{Broniowski}}, \bibinfo
  {author} {\bibfnamefont{M.}~\bibnamefont{Rybczy\'nski}},\ and\ \bibinfo
  {author} {\bibfnamefont{P.}~\bibnamefont{Bo\.zek}},\ }%
  \bibfield{journal}{%
  \Doi{10.1016/j.cpc.2008.07.016}{\bibinfo {journal} {Comput. Phys. Commun.}}\
  }%
  \textbf{\bibinfo {volume} {180}},\ \bibinfo {pages} {69} (\bibinfo {year}
  {2009})%
  \bibAnnoteFile{NoStop}{Broniowski:2007nz}%
\bibitem{Glauber:1959aa}%
  \BibitemOpen
  \bibfield{author}{%
  \bibinfo {author} {\bibfnamefont{R.~J.}\ \bibnamefont{Glauber}}\ }%
  \bibinfo {note} {~in {\it Lectures in Theoretical Physics} W.~E. Brittin and
  L.~G. Dunham eds., (Interscience, New York, 1959) Vol. 1, p. 315}%
  \bibAnnoteFile{NoStop}{Glauber:1959aa}%
\bibitem{Czyz:1969jg}%
  \BibitemOpen
  \bibfield{author}{%
  \bibinfo {author} {\bibfnamefont{W.}~\bibnamefont{Czyz}}\ and\ \bibinfo
  {author} {\bibfnamefont{L.~C.}\ \bibnamefont{Maximon}},\ }%
  \bibfield{journal}{%
  \Doi{10.1016/0003-4916(69)90321-2}{\bibinfo {journal} {Annals Phys.}}\ }%
  \textbf{\bibinfo {volume} {52}},\ \bibinfo {pages} {59} (\bibinfo {year}
  {1969})%
  \bibAnnoteFile{NoStop}{Czyz:1969jg}%
\bibitem{Bialas:2008zza}%
  \BibitemOpen
  \bibfield{author}{%
  \bibinfo {author} {\bibfnamefont{A.}~\bibnamefont{Bialas}},\ }%
  \bibfield{journal}{%
  \Doi{10.1088/0954-3899/35/4/044053}{\bibinfo {journal} {J. Phys.}}\ }%
  \textbf{\bibinfo {volume} {G35}},\ \bibinfo {pages} {044053} (\bibinfo {year}
  {2008})%
  \bibAnnoteFile{NoStop}{Bialas:2008zza}%
\bibitem{Back:2004dy}%
  \BibitemOpen
  \bibfield{author}{%
  \bibinfo {author} {\bibfnamefont{B.~B.}\ \bibnamefont{Back}} \emph{et~al.}
  (\bibinfo {collaboration} {PHOBOS}),\ }%
  \bibfield{journal}{%
  \bibinfo {journal} {Phys. Rev.}\ }%
  \textbf{\bibinfo {volume} {C70}},\ \bibinfo {pages} {021902} (\bibinfo {year}
  {2004})%
  \bibAnnoteFile{NoStop}{Back:2004dy}%
\bibitem{Bozek:2011wa}%
  \BibitemOpen
  \bibfield{author}{%
  \bibinfo {author} {\bibfnamefont{P.}~\bibnamefont{Bozek}}}%
   (\bibinfo {year} {2011}),\
  \Eprint{http://arxiv.org/abs/1101.1791}{arXiv:1101.1791 [nucl-th]}%
  \bibAnnoteFile{NoStop}{Bozek:2011wa}%
\bibitem{Wang:1991hta}%
  \BibitemOpen
  \bibfield{author}{%
  \bibinfo {author} {\bibfnamefont{X.-N.}\ \bibnamefont{Wang}}\ and\ \bibinfo
  {author} {\bibfnamefont{M.}~\bibnamefont{Gyulassy}},\ }%
  \bibfield{journal}{%
  \bibinfo {journal} {Phys. Rev.}\ }%
  \textbf{\bibinfo {volume} {D44}},\ \bibinfo {pages} {3501} (\bibinfo {year}
  {1991})%
  \bibAnnoteFile{NoStop}{Wang:1991hta}%
\bibitem{Werner:1988jr}%
  \BibitemOpen
  \bibfield{author}{%
  \bibinfo {author} {\bibfnamefont{K.}~\bibnamefont{Werner}},\ }%
  \bibfield{journal}{%
  \bibinfo {journal} {Phys. Lett.}\ }%
  \textbf{\bibinfo {volume} {B208}},\ \bibinfo {pages} {520} (\bibinfo {year}
  {1988})%
  \bibAnnoteFile{NoStop}{Werner:1988jr}%
\bibitem{Alver:2008aq}%
  \BibitemOpen
  \bibfield{author}{%
  \bibinfo {author} {\bibfnamefont{B.}~\bibnamefont{Alver}}, \bibinfo {author}
  {\bibfnamefont{M.}~\bibnamefont{Baker}}, \bibinfo {author}
  {\bibfnamefont{C.}~\bibnamefont{Loizides}},\ and\ \bibinfo {author}
  {\bibfnamefont{P.}~\bibnamefont{Steinberg}}}%
   (\bibinfo {year} {2008}),\
  \Eprint{http://arxiv.org/abs/0805.4411}{arXiv:0805.4411 [nucl-ex]}%
  \bibAnnoteFile{NoStop}{Alver:2008aq}%
\bibitem{Miller:2007ri}%
  \BibitemOpen
  \bibfield{author}{%
  \bibinfo {author} {\bibfnamefont{M.~L.}\ \bibnamefont{Miller}}, \bibinfo
  {author} {\bibfnamefont{K.}~\bibnamefont{Reygers}}, \bibinfo {author}
  {\bibfnamefont{S.~J.}\ \bibnamefont{Sanders}},\ and\ \bibinfo {author}
  {\bibfnamefont{P.}~\bibnamefont{Steinberg}},\ }%
  \bibfield{journal}{%
  \Doi{10.1146/annurev.nucl.57.090506.123020}{\bibinfo {journal} {Ann. Rev.
  Nucl. Part. Sci.}}\ }%
  \textbf{\bibinfo {volume} {57}},\ \bibinfo {pages} {205} (\bibinfo {year}
  {2007})%
  \bibAnnoteFile{NoStop}{Miller:2007ri}%
\bibitem{Chojnacki:2007jc}%
  \BibitemOpen
  \bibfield{author}{%
  \bibinfo {author} {\bibfnamefont{M.}~\bibnamefont{Chojnacki}}\ and\ \bibinfo
  {author} {\bibfnamefont{W.}~\bibnamefont{Florkowski}},\ }%
  \bibfield{journal}{%
  \bibinfo {journal} {Acta Phys. Polon.}\ }%
  \textbf{\bibinfo {volume} {B38}},\ \bibinfo {pages} {3249} (\bibinfo {year}
  {2007})%
  \bibAnnoteFile{NoStop}{Chojnacki:2007jc}%
\bibitem{Broniowski:2008vp}%
  \BibitemOpen
  \bibfield{author}{%
  \bibinfo {author} {\bibfnamefont{W.}~\bibnamefont{Broniowski}}, \bibinfo
  {author} {\bibfnamefont{M.}~\bibnamefont{Chojnacki}}, \bibinfo {author}
  {\bibfnamefont{W.}~\bibnamefont{Florkowski}},\ and\ \bibinfo {author}
  {\bibfnamefont{A.}~\bibnamefont{Kisiel}},\ }%
  \bibfield{journal}{%
  \Doi{10.1103/PhysRevLett.101.022301}{\bibinfo {journal} {Phys. Rev. Lett.}}\
  }%
  \textbf{\bibinfo {volume} {101}},\ \bibinfo {pages} {022301} (\bibinfo {year}
  {2008})%
  \bibAnnoteFile{NoStop}{Broniowski:2008vp}%
\bibitem{Huovinen:2009yb}%
  \BibitemOpen
  \bibfield{author}{%
  \bibinfo {author} {\bibfnamefont{P.}~\bibnamefont{Huovinen}}\ and\ \bibinfo
  {author} {\bibfnamefont{P.}~\bibnamefont{Petreczky}},\ }%
  \bibfield{journal}{%
  \Doi{10.1016/j.nuclphysa.2010.02.015}{\bibinfo {journal} {Nucl. Phys.}}\ }%
  \textbf{\bibinfo {volume} {A837}},\ \bibinfo {pages} {26} (\bibinfo {year}
  {2010})%
  \bibAnnoteFile{NoStop}{Huovinen:2009yb}%
\bibitem{Schenke:2010nt}%
  \BibitemOpen
  \bibfield{author}{%
  \bibinfo {author} {\bibfnamefont{B.}~\bibnamefont{Schenke}}, \bibinfo
  {author} {\bibfnamefont{S.}~\bibnamefont{Jeon}},\ and\ \bibinfo {author}
  {\bibfnamefont{C.}~\bibnamefont{Gale}},\ }%
  \bibfield{journal}{%
  \Doi{10.1103/PhysRevC.82.014903}{\bibinfo {journal} {Phys.Rev.}}\ }%
  \textbf{\bibinfo {volume} {C82}},\ \bibinfo {pages} {014903} (\bibinfo {year}
  {2010})%
  \bibAnnoteFile{NoStop}{Schenke:2010nt}%
\bibitem{Muronga:2001zk}%
  \BibitemOpen
  \bibfield{author}{%
  \bibinfo {author} {\bibfnamefont{A.}~\bibnamefont{Muronga}},\ }%
  \bibfield{journal}{%
  \Doi{10.1103/PhysRevLett.88.062302}{\bibinfo {journal} {Phys. Rev. Lett.}}\
  }%
  \textbf{\bibinfo {volume} {88}},\ \bibinfo {pages} {062302} (\bibinfo {year}
  {2002})%
  \bibAnnoteFile{NoStop}{Muronga:2001zk}%
\bibitem{Teaney:2003kp}%
  \BibitemOpen
  \bibfield{author}{%
  \bibinfo {author} {\bibfnamefont{D.}~\bibnamefont{Teaney}},\ }%
  \bibfield{journal}{%
  \bibinfo {journal} {Phys. Rev.}\ }%
  \textbf{\bibinfo {volume} {C68}},\ \bibinfo {pages} {034913} (\bibinfo {year}
  {2003})%
  \bibAnnoteFile{NoStop}{Teaney:2003kp}%
\bibitem{Baier:2006gy}%
  \BibitemOpen
  \bibfield{author}{%
  \bibinfo {author} {\bibfnamefont{R.}~\bibnamefont{Baier}}\ and\ \bibinfo
  {author} {\bibfnamefont{P.}~\bibnamefont{Romatschke}},\ }%
  \bibfield{journal}{%
  \Doi{10.1140/epjc/s10052-007-0308-5}{\bibinfo {journal} {Eur. Phys. J.}}\ }%
  \textbf{\bibinfo {volume} {C51}},\ \bibinfo {pages} {677} (\bibinfo {year}
  {2007})%
  \bibAnnoteFile{NoStop}{Baier:2006gy}%
\bibitem{Romatschke:2007mq}%
  \BibitemOpen
  \bibfield{author}{%
  \bibinfo {author} {\bibfnamefont{P.}~\bibnamefont{Romatschke}}\ and\ \bibinfo
  {author} {\bibfnamefont{U.}~\bibnamefont{Romatschke}},\ }%
  \bibfield{journal}{%
  \Doi{10.1103/PhysRevLett.99.172301}{\bibinfo {journal} {Phys. Rev. Lett.}}\
  }%
  \textbf{\bibinfo {volume} {99}},\ \bibinfo {pages} {172301} (\bibinfo {year}
  {2007})%
  \bibAnnoteFile{NoStop}{Romatschke:2007mq}%
\bibitem{Chaudhuri:2006jd}%
  \BibitemOpen
  \bibfield{author}{%
  \bibinfo {author} {\bibfnamefont{A.~K.}\ \bibnamefont{Chaudhuri}},\ }%
  \bibfield{journal}{%
  \bibinfo {journal} {Phys. Rev.}\ }%
  \textbf{\bibinfo {volume} {C74}},\ \bibinfo {pages} {044904} (\bibinfo {year}
  {2006})%
  \bibAnnoteFile{NoStop}{Chaudhuri:2006jd}%
\bibitem{Song:2007fn}%
  \BibitemOpen
  \bibfield{author}{%
  \bibinfo {author} {\bibfnamefont{H.}~\bibnamefont{Song}}\ and\ \bibinfo
  {author} {\bibfnamefont{U.~W.}\ \bibnamefont{Heinz}},\ }%
  \bibfield{journal}{%
  \Doi{10.1016/j.physletb.2007.11.019}{\bibinfo {journal} {Phys. Lett.}}\ }%
  \textbf{\bibinfo {volume} {B658}},\ \bibinfo {pages} {279} (\bibinfo {year}
  {2008})%
  \bibAnnoteFile{NoStop}{Song:2007fn}%
\bibitem{Bozek:2007qt}%
  \BibitemOpen
  \bibfield{author}{%
  \bibinfo {author} {\bibfnamefont{P.}~\bibnamefont{Bo\.zek}},\ }%
  \bibfield{journal}{%
  \Doi{10.1103/PhysRevC.77.034911}{\bibinfo {journal} {Phys. Rev.}}\ }%
  \textbf{\bibinfo {volume} {C77}},\ \bibinfo {pages} {034911} (\bibinfo {year}
  {2008})%
  \bibAnnoteFile{NoStop}{Bozek:2007qt}%
\bibitem{Bozek:2009dw}%
  \BibitemOpen
  \bibfield{author}{%
  \bibinfo {author} {\bibfnamefont{P.}~\bibnamefont{Bo\.zek}},\ }%
  \bibfield{journal}{%
  \bibinfo {journal} {Phys. Rev.}\ }%
  \textbf{\bibinfo {volume} {C81}},\ \bibinfo {pages} {034909} (\bibinfo {year}
  {2010})%
  \bibAnnoteFile{NoStop}{Bozek:2009dw}%
\bibitem{Schenke:2010rr}%
  \BibitemOpen
  \bibfield{author}{%
  \bibinfo {author} {\bibfnamefont{B.}~\bibnamefont{Schenke}}, \bibinfo
  {author} {\bibfnamefont{S.}~\bibnamefont{Jeon}},\ and\ \bibinfo {author}
  {\bibfnamefont{C.}~\bibnamefont{Gale}},\ }%
  \bibfield{journal}{%
  \Doi{10.1103/PhysRevLett.106.042301}{\bibinfo {journal} {Phys.Rev.Lett.}}\ }%
  \textbf{\bibinfo {volume} {106}},\ \bibinfo {pages} {042301} (\bibinfo {year}
  {2011})%
  \bibAnnoteFile{NoStop}{Schenke:2010rr}%
\bibitem{Schenke:2011tv}%
  \BibitemOpen
  \bibfield{author}{%
  \bibinfo {author} {\bibfnamefont{B.}~\bibnamefont{Schenke}}, \bibinfo
  {author} {\bibfnamefont{S.}~\bibnamefont{Jeon}},\ and\ \bibinfo {author}
  {\bibfnamefont{C.}~\bibnamefont{Gale}},\ }%
  \bibfield{journal}{%
  \Doi{10.1016/j.physletb.2011.06.065}{\bibinfo {journal} {Phys.Lett.}}\ }%
  \textbf{\bibinfo {volume} {B702}},\ \bibinfo {pages} {59} (\bibinfo {year}
  {2011})%
  \bibAnnoteFile{NoStop}{Schenke:2011tv}%
\bibitem{Aguiar:2000hw}%
  \BibitemOpen
  \bibfield{author}{%
  \bibinfo {author} {\bibfnamefont{C.~E.}\ \bibnamefont{Aguiar}}, \bibinfo
  {author} {\bibfnamefont{T.}~\bibnamefont{Kodama}}, \bibinfo {author}
  {\bibfnamefont{T.}~\bibnamefont{Osada}},\ and\ \bibinfo {author}
  {\bibfnamefont{Y.}~\bibnamefont{Hama}},\ }%
  \bibfield{journal}{%
  \bibinfo {journal} {J. Phys.}\ }%
  \textbf{\bibinfo {volume} {G27}},\ \bibinfo {pages} {75} (\bibinfo {year}
  {2001})%
  \bibAnnoteFile{NoStop}{Aguiar:2000hw}%
\bibitem{Miller:2003kd}%
  \BibitemOpen
  \bibfield{author}{%
  \bibinfo {author} {\bibfnamefont{M.}~\bibnamefont{Miller}}\ and\ \bibinfo
  {author} {\bibfnamefont{R.}~\bibnamefont{Snellings}}}%
   (\bibinfo {year} {2003}),\
  \Eprint{http://arxiv.org/abs/nucl-ex/0312008}{arXiv:nucl-ex/0312008}%
  \bibAnnoteFile{NoStop}{Miller:2003kd}%
\bibitem{Bhalerao:2005mm}%
  \BibitemOpen
  \bibfield{author}{%
  \bibinfo {author} {\bibfnamefont{R.~S.}\ \bibnamefont{Bhalerao}}, \bibinfo
  {author} {\bibfnamefont{J.-P.}\ \bibnamefont{Blaizot}}, \bibinfo {author}
  {\bibfnamefont{N.}~\bibnamefont{Borghini}},\ and\ \bibinfo {author}
  {\bibfnamefont{J.-Y.}\ \bibnamefont{Ollitrault}},\ }%
  \bibfield{journal}{%
  \bibinfo {journal} {Phys. Lett.}\ }%
  \textbf{\bibinfo {volume} {B627}},\ \bibinfo {pages} {49} (\bibinfo {year}
  {2005})%
  \bibAnnoteFile{NoStop}{Bhalerao:2005mm}%
\bibitem{Manly:2005zy}%
  \BibitemOpen
  \bibfield{author}{%
  \bibinfo {author} {\bibfnamefont{S.}~\bibnamefont{Manly}} \emph{et~al.}
  (\bibinfo {collaboration} {PHOBOS}),\ }%
  \bibfield{journal}{%
  \Doi{10.1016/j.nuclphysa.2006.06.079}{\bibinfo {journal} {Nucl. Phys.}}\ }%
  \textbf{\bibinfo {volume} {A774}},\ \bibinfo {pages} {523} (\bibinfo {year}
  {2006})%
  \bibAnnoteFile{NoStop}{Manly:2005zy}%
\bibitem{Andrade:2006yh}%
  \BibitemOpen
  \bibfield{author}{%
  \bibinfo {author} {\bibfnamefont{R.}~\bibnamefont{Andrade}}, \bibinfo
  {author} {\bibfnamefont{F.}~\bibnamefont{Grassi}}, \bibinfo {author}
  {\bibfnamefont{Y.}~\bibnamefont{Hama}}, \bibinfo {author}
  {\bibfnamefont{T.}~\bibnamefont{Kodama}},\ and\ \bibinfo {author}
  {\bibfnamefont{J.}~\bibnamefont{Socolowski}, \bibfnamefont{O.}},\ }%
  \bibfield{journal}{%
  \bibinfo {journal} {Phys. Rev. Lett.}\ }%
  \textbf{\bibinfo {volume} {97}},\ \bibinfo {pages} {202302} (\bibinfo {year}
  {2006})%
  \bibAnnoteFile{NoStop}{Andrade:2006yh}%
\bibitem{Voloshin:2006gz}%
  \BibitemOpen
  \bibfield{author}{%
  \bibinfo {author} {\bibfnamefont{S.~A.}\ \bibnamefont{Voloshin}}}%
   (\bibinfo {year} {2006}),\
  \Eprint{http://arxiv.org/abs/nucl-th/0606022}{arXiv:nucl-th/0606022}%
  \bibAnnoteFile{NoStop}{Voloshin:2006gz}%
\bibitem{Alver:2006pn}%
  \BibitemOpen
  \bibfield{author}{%
  \bibinfo {author} {\bibfnamefont{B.}~\bibnamefont{Alver}} \emph{et~al.}
  (\bibinfo {collaboration} {PHOBOS}),\ }%
  \bibfield{journal}{%
  \bibinfo {journal} {PoS}\ }%
  \textbf{\bibinfo {volume} {CFRNC2006}},\ \bibinfo {pages} {023} (\bibinfo
  {year} {2006})%
  \bibAnnoteFile{NoStop}{Alver:2006pn}%
\bibitem{Drescher:2006ca}%
  \BibitemOpen
  \bibfield{author}{%
  \bibinfo {author} {\bibfnamefont{H.~J.}\ \bibnamefont{Drescher}}\ and\
  \bibinfo {author} {\bibfnamefont{Y.}~\bibnamefont{Nara}},\ }%
  \bibfield{journal}{%
  \bibinfo {journal} {Phys. Rev.}\ }%
  \textbf{\bibinfo {volume} {C75}},\ \bibinfo {pages} {034905} (\bibinfo {year}
  {2007})%
  \bibAnnoteFile{NoStop}{Drescher:2006ca}%
\bibitem{Alver:2006wh}%
  \BibitemOpen
  \bibfield{author}{%
  \bibinfo {author} {\bibfnamefont{B.}~\bibnamefont{Alver}} \emph{et~al.}
  (\bibinfo {collaboration} {PHOBOS})}%
   (\bibinfo {year} {2006}),\
  \Eprint{http://arxiv.org/abs/nucl-ex/0610037}{arXiv:nucl-ex/0610037}%
  \bibAnnoteFile{NoStop}{Alver:2006wh}%
\bibitem{Alver:2006zz}%
  \BibitemOpen
  \bibfield{author}{%
  \bibinfo {author} {\bibfnamefont{B.}~\bibnamefont{Alver}} \emph{et~al.}
  (\bibinfo {collaboration} {PHOBOS Collaboration}),\ }%
  \bibfield{journal}{%
  \bibinfo {journal} {PoS}\ }%
  \textbf{\bibinfo {volume} {CFRNC2006}},\ \bibinfo {pages} {023} (\bibinfo
  {year} {2006})%
  \bibAnnoteFile{NoStop}{Alver:2006zz}%
\bibitem{Sorensen:2006nw}%
  \BibitemOpen
  \bibfield{author}{%
  \bibinfo {author} {\bibfnamefont{P.}~\bibnamefont{Sorensen}} (\bibinfo
  {collaboration} {STAR})}%
   (\bibinfo {year} {2006}),\
  \Eprint{http://arxiv.org/abs/nucl-ex/0612021}{nucl-ex/0612021}%
  \bibAnnoteFile{NoStop}{Sorensen:2006nw}%
\bibitem{Broniowski:2007ft}%
  \BibitemOpen
  \bibfield{author}{%
  \bibinfo {author} {\bibfnamefont{W.}~\bibnamefont{Broniowski}}, \bibinfo
  {author} {\bibfnamefont{P.}~\bibnamefont{Bo\.zek}},\ and\ \bibinfo {author}
  {\bibfnamefont{M.}~\bibnamefont{Rybczy\'nski}},\ }%
  \bibfield{journal}{%
  \bibinfo {journal} {Phys. Rev.}\ }%
  \textbf{\bibinfo {volume} {C76}},\ \bibinfo {pages} {054905} (\bibinfo {year}
  {2007})%
  \bibAnnoteFile{NoStop}{Broniowski:2007ft}%
\bibitem{Alver:2007rm}%
  \BibitemOpen
  \bibfield{author}{%
  \bibinfo {author} {\bibfnamefont{B.}~\bibnamefont{Alver}} \emph{et~al.}
  (\bibinfo {collaboration} {PHOBOS}),\ }%
  \bibfield{journal}{%
  \Doi{10.1088/0954-3899/34/8/S123}{\bibinfo {journal} {J. Phys.}}\ }%
  \textbf{\bibinfo {volume} {G34}},\ \bibinfo {pages} {S907} (\bibinfo {year}
  {2007})%
  \bibAnnoteFile{NoStop}{Alver:2007rm}%
\bibitem{Hama:2007dq}%
  \BibitemOpen
  \bibfield{author}{%
  \bibinfo {author} {\bibfnamefont{Y.}~\bibnamefont{Hama}} \emph{et~al.}}%
   (\bibinfo {year} {2007}),\
  \Eprint{http://arxiv.org/abs/0711.4544}{arXiv:0711.4544 [hep-ph]}%
  \bibAnnoteFile{NoStop}{Hama:2007dq}%
\bibitem{Voloshin:2007pc}%
  \BibitemOpen
  \bibfield{author}{%
  \bibinfo {author} {\bibfnamefont{S.~A.}\ \bibnamefont{Voloshin}}, \bibinfo
  {author} {\bibfnamefont{A.~M.}\ \bibnamefont{Poskanzer}}, \bibinfo {author}
  {\bibfnamefont{A.}~\bibnamefont{Tang}},\ and\ \bibinfo {author}
  {\bibfnamefont{G.}~\bibnamefont{Wang}}}%
   (\bibinfo {year} {2007}),\ \Eprint{http://arxiv.org/abs/arXiv:0708.0800
  [nucl-th]}{arXiv:0708.0800 [nucl-th]}%
  \bibAnnoteFile{NoStop}{Voloshin:2007pc}%
\bibitem{Andrade:2008fa}%
  \BibitemOpen
  \bibfield{author}{%
  \bibinfo {author} {\bibfnamefont{R.~P.~G.}\ \bibnamefont{Andrade}}
  \emph{et~al.},\ }%
  \bibfield{journal}{%
  \bibinfo {journal} {Acta Phys. Polon.}\ }%
  \textbf{\bibinfo {volume} {B40}},\ \bibinfo {pages} {993} (\bibinfo {year}
  {2009})%
  \bibAnnoteFile{NoStop}{Andrade:2008fa}%
\bibitem{Hama:2009pk}%
  \BibitemOpen
  \bibfield{author}{%
  \bibinfo {author} {\bibfnamefont{Y.}~\bibnamefont{Hama}}, \bibinfo {author}
  {\bibfnamefont{R.~P.~G.}\ \bibnamefont{Andrade}}, \bibinfo {author}
  {\bibfnamefont{F.}~\bibnamefont{Grassi}}, \bibinfo {author}
  {\bibfnamefont{W.~L.}\ \bibnamefont{Qian}},\ and\ \bibinfo {author}
  {\bibfnamefont{T.}~\bibnamefont{Kodama}},\ }%
  \bibfield{journal}{%
  \bibinfo {journal} {Acta Phys. Polon.}\ }%
  \textbf{\bibinfo {volume} {B40}},\ \bibinfo {pages} {931} (\bibinfo {year}
  {2009})%
  \bibAnnoteFile{NoStop}{Hama:2009pk}%
\bibitem{Alt:2006jr}%
  \BibitemOpen
  \bibfield{author}{%
  \bibinfo {author} {\bibfnamefont{C.}~\bibnamefont{Alt}} \emph{et~al.}
  (\bibinfo {collaboration} {NA49}),\ }%
  \bibfield{journal}{%
  \Doi{10.1103/PhysRevC.75.064904}{\bibinfo {journal} {Phys. Rev.}}\ }%
  \textbf{\bibinfo {volume} {C75}},\ \bibinfo {pages} {064904} (\bibinfo {year}
  {2007})%
  \bibAnnoteFile{NoStop}{Alt:2006jr}%
\bibitem{Broniowski:2009fm}%
  \BibitemOpen
  \bibfield{author}{%
  \bibinfo {author} {\bibfnamefont{W.}~\bibnamefont{Broniowski}}, \bibinfo
  {author} {\bibfnamefont{M.}~\bibnamefont{Chojnacki}},\ and\ \bibinfo {author}
  {\bibfnamefont{L.}~\bibnamefont{Obara}},\ }%
  \bibfield{journal}{%
  \Doi{10.1103/PhysRevC.80.051902}{\bibinfo {journal} {Phys. Rev.}}\ }%
  \textbf{\bibinfo {volume} {C80}},\ \bibinfo {pages} {051902} (\bibinfo {year}
  {2009})%
  \bibAnnoteFile{NoStop}{Broniowski:2009fm}%
\bibitem{Gazdzicki:1992ri}%
  \BibitemOpen
  \bibfield{author}{%
  \bibinfo {author} {\bibfnamefont{M.}~\bibnamefont{Gazdzicki}}\ and\ \bibinfo
  {author} {\bibfnamefont{S.}~\bibnamefont{Mrowczynski}},\ }%
  \bibfield{journal}{%
  \bibinfo {journal} {Z. Phys.}\ }%
  \textbf{\bibinfo {volume} {C54}},\ \bibinfo {pages} {127} (\bibinfo {year}
  {1992})%
  \bibAnnoteFile{NoStop}{Gazdzicki:1992ri}%
\bibitem{Stodolsky:1995ds}%
  \BibitemOpen
  \bibfield{author}{%
  \bibinfo {author} {\bibfnamefont{L.}~\bibnamefont{Stodolsky}},\ }%
  \bibfield{journal}{%
  \bibinfo {journal} {Phys. Rev. Lett.}\ }%
  \textbf{\bibinfo {volume} {75}},\ \bibinfo {pages} {1044} (\bibinfo {year}
  {1995})%
  \bibAnnoteFile{NoStop}{Stodolsky:1995ds}%
\bibitem{Shuryak:1997yj}%
  \BibitemOpen
  \bibfield{author}{%
  \bibinfo {author} {\bibfnamefont{E.~V.}\ \bibnamefont{Shuryak}},\ }%
  \bibfield{journal}{%
  \bibinfo {journal} {Phys. Lett.}\ }%
  \textbf{\bibinfo {volume} {B423}},\ \bibinfo {pages} {9} (\bibinfo {year}
  {1998})%
  \bibAnnoteFile{NoStop}{Shuryak:1997yj}%
\bibitem{Mrowczynski:1997kz}%
  \BibitemOpen
  \bibfield{author}{%
  \bibinfo {author} {\bibfnamefont{S.}~\bibnamefont{Mrowczynski}},\ }%
  \bibfield{journal}{%
  \bibinfo {journal} {Phys. Lett.}\ }%
  \textbf{\bibinfo {volume} {B430}},\ \bibinfo {pages} {9} (\bibinfo {year}
  {1998})%
  \bibAnnoteFile{NoStop}{Mrowczynski:1997kz}%
\bibitem{Voloshin:1999yf}%
  \BibitemOpen
  \bibfield{author}{%
  \bibinfo {author} {\bibfnamefont{S.~A.}\ \bibnamefont{Voloshin}}, \bibinfo
  {author} {\bibfnamefont{V.}~\bibnamefont{Koch}},\ and\ \bibinfo {author}
  {\bibfnamefont{H.~G.}\ \bibnamefont{Ritter}},\ }%
  \bibfield{journal}{%
  \bibinfo {journal} {Phys. Rev.}\ }%
  \textbf{\bibinfo {volume} {C60}},\ \bibinfo {pages} {024901} (\bibinfo {year}
  {1999})%
  \bibAnnoteFile{NoStop}{Voloshin:1999yf}%
\bibitem{Baym:1999up}%
  \BibitemOpen
  \bibfield{author}{%
  \bibinfo {author} {\bibfnamefont{G.}~\bibnamefont{Baym}}\ and\ \bibinfo
  {author} {\bibfnamefont{H.}~\bibnamefont{Heiselberg}},\ }%
  \bibfield{journal}{%
  \bibinfo {journal} {Phys. Lett.}\ }%
  \textbf{\bibinfo {volume} {B469}},\ \bibinfo {pages} {7} (\bibinfo {year}
  {1999})%
  \bibAnnoteFile{NoStop}{Baym:1999up}%
\bibitem{Appelshauser:1999ft}%
  \BibitemOpen
  \bibfield{author}{%
  \bibinfo {author} {\bibfnamefont{H.}~\bibnamefont{Appelshauser}}
  \emph{et~al.} (\bibinfo {collaboration} {NA49}),\ }%
  \bibfield{journal}{%
  \bibinfo {journal} {Phys. Lett.}\ }%
  \textbf{\bibinfo {volume} {B459}},\ \bibinfo {pages} {679} (\bibinfo {year}
  {1999})%
  \bibAnnoteFile{NoStop}{Appelshauser:1999ft}%
\bibitem{Voloshin:2001ei}%
  \BibitemOpen
  \bibfield{author}{%
  \bibinfo {author} {\bibfnamefont{S.~A.}\ \bibnamefont{Voloshin}} (\bibinfo
  {collaboration} {STAR})}%
   (\bibinfo {year} {2001}),\
  \Eprint{http://arxiv.org/abs/nucl-ex/0109006}{arXiv:nucl-ex/0109006}%
  \bibAnnoteFile{NoStop}{Voloshin:2001ei}%
\bibitem{Prindle:2006zz}%
  \BibitemOpen
  \bibfield{author}{%
  \bibinfo {author} {\bibfnamefont{D.~J.}\ \bibnamefont{Prindle}}\ and\
  \bibinfo {author} {\bibfnamefont{T.~A.}\ \bibnamefont{Trainor}} (\bibinfo
  {collaboration} {STAR}),\ }%
  \bibfield{journal}{%
  \bibinfo {journal} {PoS}\ }%
  \textbf{\bibinfo {volume} {CFRNC2006}},\ \bibinfo {pages} {007} (\bibinfo
  {year} {2006})%
  \bibAnnoteFile{NoStop}{Prindle:2006zz}%
\bibitem{Mrowczynski:2009wk}%
  \BibitemOpen
  \bibfield{author}{%
  \bibinfo {author} {\bibfnamefont{S.}~\bibnamefont{Mrowczynski}},\ }%
  \bibfield{journal}{%
  \bibinfo {journal} {Acta Phys. Polon.}\ }%
  \textbf{\bibinfo {volume} {B40}},\ \bibinfo {pages} {1053} (\bibinfo {year}
  {2009})%
  \bibAnnoteFile{NoStop}{Mrowczynski:2009wk}%
\bibitem{Adams:2003uw}%
  \BibitemOpen
  \bibfield{author}{%
  \bibinfo {author} {\bibfnamefont{J.}~\bibnamefont{Adams}} \emph{et~al.}
  (\bibinfo {collaboration} {STAR}),\ }%
  \bibfield{journal}{%
  \bibinfo {journal} {Phys. Rev.}\ }%
  \textbf{\bibinfo {volume} {C71}},\ \bibinfo {pages} {064906} (\bibinfo {year}
  {2005})%
  \bibAnnoteFile{NoStop}{Adams:2003uw}%
\bibitem{Adamova:2003pz}%
  \BibitemOpen
  \bibfield{author}{%
  \bibinfo {author} {\bibfnamefont{D.}~\bibnamefont{Adamova}} \emph{et~al.}
  (\bibinfo {collaboration} {CERES}),\ }%
  \bibfield{journal}{%
  \Doi{10.1016/j.nuclphysa.2003.07.018}{\bibinfo {journal} {Nucl. Phys.}}\ }%
  \textbf{\bibinfo {volume} {A727}},\ \bibinfo {pages} {97} (\bibinfo {year}
  {2003})%
  \bibAnnoteFile{NoStop}{Adamova:2003pz}%
\bibitem{Adler:2003xq}%
  \BibitemOpen
  \bibfield{author}{%
  \bibinfo {author} {\bibfnamefont{S.~S.}\ \bibnamefont{Adler}} \emph{et~al.}
  (\bibinfo {collaboration} {PHENIX}),\ }%
  \bibfield{journal}{%
  \bibinfo {journal} {Phys. Rev. Lett.}\ }%
  \textbf{\bibinfo {volume} {93}},\ \bibinfo {pages} {092301} (\bibinfo {year}
  {2004})%
  \bibAnnoteFile{NoStop}{Adler:2003xq}%
\bibitem{Adams:2005ka}%
  \BibitemOpen
  \bibfield{author}{%
  \bibinfo {author} {\bibfnamefont{J.}~\bibnamefont{Adams}} \emph{et~al.}
  (\bibinfo {collaboration} {STAR})}%
   (\bibinfo {year} {2005}),\
  \Eprint{http://arxiv.org/abs/nucl-ex/0504031}{nucl-ex/0504031}%
  \bibAnnoteFile{NoStop}{Adams:2005ka}%
\bibitem{Grebieszkow:2007xz}%
  \BibitemOpen
  \bibfield{author}{%
  \bibinfo {author} {\bibfnamefont{K.}~\bibnamefont{Grebieszkow}}
  \emph{et~al.},\ }%
  \bibfield{journal}{%
  \bibinfo {journal} {PoS}\ }%
  \textbf{\bibinfo {volume} {CPOD07}},\ \bibinfo {pages} {022} (\bibinfo {year}
  {2007})%
  \bibAnnoteFile{NoStop}{Grebieszkow:2007xz}%
\bibitem{na49:2008vb}%
  \BibitemOpen
  \bibfield{author}{%
  \bibinfo {author} {\bibfnamefont{T.}~\bibnamefont{Anticic}} \emph{et~al.}
  (\bibinfo {collaboration} {NA49}),\ }%
  \bibfield{journal}{%
  \bibinfo {journal} {Phys. Rev.}\ }%
  \textbf{\bibinfo {volume} {C79}},\ \bibinfo {pages} {044904} (\bibinfo {year}
  {2009})%
  \bibAnnoteFile{NoStop}{na49:2008vb}%
\bibitem{Bialas:2004su}%
  \BibitemOpen
  \bibfield{author}{%
  \bibinfo {author} {\bibfnamefont{A.}~\bibnamefont{Bia\l{}as}}\ and\ \bibinfo
  {author} {\bibfnamefont{W.}~\bibnamefont{Czy\.z}},\ }%
  \bibfield{journal}{%
  \bibinfo {journal} {Acta Phys. Polon.}\ }%
  \textbf{\bibinfo {volume} {B36}},\ \bibinfo {pages} {905} (\bibinfo {year}
  {2005})%
  \bibAnnoteFile{NoStop}{Bialas:2004su}%
\bibitem{Gazdzicki:2005rr}%
  \BibitemOpen
  \bibfield{author}{%
  \bibinfo {author} {\bibfnamefont{M.}~\bibnamefont{Ga\'zdzicki}}\ and\
  \bibinfo {author} {\bibfnamefont{M.~I.}\ \bibnamefont{Gorenstein}},\ }%
  \bibfield{journal}{%
  \Doi{10.1016/j.physletb.2006.07.044}{\bibinfo {journal} {Phys. Lett.}}\ }%
  \textbf{\bibinfo {volume} {B640}},\ \bibinfo {pages} {155} (\bibinfo {year}
  {2006})%
  \bibAnnoteFile{NoStop}{Gazdzicki:2005rr}%
\bibitem{Bzdak:2009dr}%
  \BibitemOpen
  \bibfield{author}{%
  \bibinfo {author} {\bibfnamefont{A.}~\bibnamefont{Bzdak}}\ and\ \bibinfo
  {author} {\bibfnamefont{K.}~\bibnamefont{Wozniak}},\ }%
  \bibfield{journal}{%
  \Doi{10.1103/PhysRevC.81.034908}{\bibinfo {journal} {Phys. Rev.}}\ }%
  \textbf{\bibinfo {volume} {C81}},\ \bibinfo {pages} {034908} (\bibinfo {year}
  {2010})%
  \bibAnnoteFile{NoStop}{Bzdak:2009dr}%
\bibitem{Bzdak:2009xq}%
  \BibitemOpen
  \bibfield{author}{%
  \bibinfo {author} {\bibfnamefont{A.}~\bibnamefont{Bzdak}},\ }%
  \bibfield{journal}{%
  \Doi{10.1103/PhysRevC.80.024906}{\bibinfo {journal} {Phys. Rev.}}\ }%
  \textbf{\bibinfo {volume} {C80}},\ \bibinfo {pages} {024906} (\bibinfo {year}
  {2009})%
  \bibAnnoteFile{NoStop}{Bzdak:2009xq}%
\bibitem{Bzdak:2009zz}%
  \BibitemOpen
  \bibfield{author}{%
  \bibinfo {author} {\bibfnamefont{A.}~\bibnamefont{Bzdak}},\ }%
  \bibfield{journal}{%
  \bibinfo {journal} {Acta Phys. Polon.}\ }%
  \textbf{\bibinfo {volume} {B40}},\ \bibinfo {pages} {2029} (\bibinfo {year}
  {2009})%
  \bibAnnoteFile{NoStop}{Bzdak:2009zz}%
\bibitem{Bialas:2010zb}%
  \BibitemOpen
  \bibfield{author}{%
  \bibinfo {author} {\bibfnamefont{A.}~\bibnamefont{Bialas}}\ and\ \bibinfo
  {author} {\bibfnamefont{K.}~\bibnamefont{Zalewski}},\ }%
  \bibfield{journal}{%
  \Doi{10.1103/PhysRevC.82.034911}{\bibinfo {journal} {Phys. Rev.}}\ }%
  \textbf{\bibinfo {volume} {C82}},\ \bibinfo {pages} {034911} (\bibinfo {year}
  {2010})%
  \bibAnnoteFile{NoStop}{Bialas:2010zb}%
\bibitem{Bozek:2010bi}%
  \BibitemOpen
  \bibfield{author}{%
  \bibinfo {author} {\bibfnamefont{P.}~\bibnamefont{Bozek}}\ and\ \bibinfo
  {author} {\bibfnamefont{I.}~\bibnamefont{Wyskiel}},\ }%
  \bibfield{journal}{%
  \Doi{10.1103/PhysRevC.81.054902}{\bibinfo {journal} {Phys. Rev.}}\ }%
  \textbf{\bibinfo {volume} {C81}},\ \bibinfo {pages} {054902} (\bibinfo {year}
  {2010})%
  \bibAnnoteFile{NoStop}{Bozek:2010bi}%
\bibitem{Bozek:2010vz}%
  \BibitemOpen
  \bibfield{author}{%
  \bibinfo {author} {\bibfnamefont{P.}~\bibnamefont{Bozek}}, \bibinfo {author}
  {\bibfnamefont{W.}~\bibnamefont{Broniowski}},\ and\ \bibinfo {author}
  {\bibfnamefont{J.}~\bibnamefont{Moreira}},\ }%
  \bibfield{journal}{%
  \Doi{10.1103/PhysRevC.83.034911}{\bibinfo {journal} {Phys.Rev.}}\ }%
  \textbf{\bibinfo {volume} {C83}},\ \bibinfo {pages} {034911} (\bibinfo {year}
  {2011})%
  \bibAnnoteFile{NoStop}{Bozek:2010vz}%
\bibitem{Alver:2010gr}%
  \BibitemOpen
  \bibfield{author}{%
  \bibinfo {author} {\bibfnamefont{B.}~\bibnamefont{Alver}}\ and\ \bibinfo
  {author} {\bibfnamefont{G.}~\bibnamefont{Roland}},\ }%
  \bibfield{journal}{%
  \Doi{10.1103/PhysRevC.81.054905}{\bibinfo {journal} {Phys. Rev.}}\ }%
  \textbf{\bibinfo {volume} {C81}},\ \bibinfo {pages} {054905} (\bibinfo {year}
  {2010})%
  \bibAnnoteFile{NoStop}{Alver:2010gr}%
\bibitem{Alver:2010dn}%
  \BibitemOpen
  \bibfield{author}{%
  \bibinfo {author} {\bibfnamefont{B.~H.}\ \bibnamefont{Alver}}, \bibinfo
  {author} {\bibfnamefont{C.}~\bibnamefont{Gombeaud}}, \bibinfo {author}
  {\bibfnamefont{M.}~\bibnamefont{Luzum}},\ and\ \bibinfo {author}
  {\bibfnamefont{J.-Y.}\ \bibnamefont{Ollitrault}},\ }%
  \bibfield{journal}{%
  \Doi{10.1103/PhysRevC.82.034913}{\bibinfo {journal} {Phys. Rev.}}\ }%
  \textbf{\bibinfo {volume} {C82}},\ \bibinfo {pages} {034913} (\bibinfo {year}
  {2010})%
  \bibAnnoteFile{NoStop}{Alver:2010dn}%
\bibitem{Petersen:2010cw}%
  \BibitemOpen
  \bibfield{author}{%
  \bibinfo {author} {\bibfnamefont{H.}~\bibnamefont{Petersen}}, \bibinfo
  {author} {\bibfnamefont{G.-Y.}\ \bibnamefont{Qin}}, \bibinfo {author}
  {\bibfnamefont{S.~A.}\ \bibnamefont{Bass}},\ and\ \bibinfo {author}
  {\bibfnamefont{B.}~\bibnamefont{Muller}}}%
   (\bibinfo {year} {2010}),\
  \Eprint{http://arxiv.org/abs/1008.0625}{arXiv:1008.0625 [nucl-th]}%
  \bibAnnoteFile{NoStop}{Petersen:2010cw}%
\bibitem{Bialas:2006qf}%
  \BibitemOpen
  \bibfield{author}{%
  \bibinfo {author} {\bibfnamefont{A.}~\bibnamefont{Bia\l{}as}}\ and\ \bibinfo
  {author} {\bibfnamefont{A.}~\bibnamefont{Bzdak}},\ }%
  \bibfield{journal}{%
  \bibinfo {journal} {Acta Phys. Polon.}\ }%
  \textbf{\bibinfo {volume} {B38}},\ \bibinfo {pages} {159} (\bibinfo {year}
  {2007})%
  \bibAnnoteFile{NoStop}{Bialas:2006qf}%
\bibitem{Bohm:1974tv}%
  \BibitemOpen
  \bibfield{author}{%
  \bibinfo {author} {\bibfnamefont{A.}~\bibnamefont{Bohm}} \emph{et~al.},\ }%
  \bibfield{journal}{%
  \Doi{10.1016/0370-2693(74)90644-3}{\bibinfo {journal} {Phys. Lett.}}\ }%
  \textbf{\bibinfo {volume} {B49}},\ \bibinfo {pages} {491} (\bibinfo {year}
  {1974})%
  \bibAnnoteFile{NoStop}{Bohm:1974tv}%
\bibitem{Nagy:1978iw}%
  \BibitemOpen
  \bibfield{author}{%
  \bibinfo {author} {\bibfnamefont{E.}~\bibnamefont{Nagy}} \emph{et~al.},\ }%
  \bibfield{journal}{%
  \Doi{10.1016/0550-3213(79)90301-8}{\bibinfo {journal} {Nucl. Phys.}}\ }%
  \textbf{\bibinfo {volume} {B150}},\ \bibinfo {pages} {221} (\bibinfo {year}
  {1979})%
  \bibAnnoteFile{NoStop}{Nagy:1978iw}%
\bibitem{Amaldi:1979kd}%
  \BibitemOpen
  \bibfield{author}{%
  \bibinfo {author} {\bibfnamefont{U.}~\bibnamefont{Amaldi}}\ and\ \bibinfo
  {author} {\bibfnamefont{K.~R.}\ \bibnamefont{Schubert}},\ }%
  \bibfield{journal}{%
  \Doi{10.1016/0550-3213(80)90229-1}{\bibinfo {journal} {Nucl. Phys.}}\ }%
  \textbf{\bibinfo {volume} {B166}},\ \bibinfo {pages} {301} (\bibinfo {year}
  {1980})%
  \bibAnnoteFile{NoStop}{Amaldi:1979kd}%
\bibitem{Amos:1985wx}%
  \BibitemOpen
  \bibfield{author}{%
  \bibinfo {author} {\bibfnamefont{N.~A.}\ \bibnamefont{Amos}} \emph{et~al.},\
  }%
  \bibfield{journal}{%
  \Doi{10.1016/0550-3213(85)90511-5}{\bibinfo {journal} {Nucl. Phys.}}\ }%
  \textbf{\bibinfo {volume} {B262}},\ \bibinfo {pages} {689} (\bibinfo {year}
  {1985})%
  \bibAnnoteFile{NoStop}{Amos:1985wx}%
\bibitem{Breakstone:1984te}%
  \BibitemOpen
  \bibfield{author}{%
  \bibinfo {author} {\bibfnamefont{A.}~\bibnamefont{Breakstone}} \emph{et~al.}
  (\bibinfo {collaboration} {AMES-BOLOGNA-CERN-DORTMUND-HEIDELBERG-WARSAW}),\
  }%
  \bibfield{journal}{%
  \Doi{10.1016/0550-3213(84)90595-9}{\bibinfo {journal} {Nucl. Phys.}}\ }%
  \textbf{\bibinfo {volume} {B248}},\ \bibinfo {pages} {253} (\bibinfo {year}
  {1984})%
  \bibAnnoteFile{NoStop}{Breakstone:1984te}%
\bibitem{Broniowski:2010jd}%
  \BibitemOpen
  \bibfield{author}{%
  \bibinfo {author} {\bibfnamefont{W.}~\bibnamefont{Broniowski}}\ and\ \bibinfo
  {author} {\bibfnamefont{M.}~\bibnamefont{Rybczynski}},\ }%
  \bibfield{journal}{%
  \Doi{10.1103/PhysRevC.81.064909}{\bibinfo {journal} {Phys. Rev.}}\ }%
  \textbf{\bibinfo {volume} {C81}},\ \bibinfo {pages} {064909} (\bibinfo {year}
  {2010})%
  \bibAnnoteFile{NoStop}{Broniowski:2010jd}%
\bibitem{Alvioli:2009ab}%
  \BibitemOpen
  \bibfield{author}{%
  \bibinfo {author} {\bibfnamefont{M.}~\bibnamefont{Alvioli}}, \bibinfo
  {author} {\bibfnamefont{H.~J.}\ \bibnamefont{Drescher}},\ and\ \bibinfo
  {author} {\bibfnamefont{M.}~\bibnamefont{Strikman}},\ }%
  \bibfield{journal}{%
  \Doi{10.1016/j.physletb.2009.08.067}{\bibinfo {journal} {Phys. Lett.}}\ }%
  \textbf{\bibinfo {volume} {B680}},\ \bibinfo {pages} {225} (\bibinfo {year}
  {2009})%
  \bibAnnoteFile{NoStop}{Alvioli:2009ab}%
\bibitem{Alvioli:2010yk}%
  \BibitemOpen
  \bibfield{author}{%
  \bibinfo {author} {\bibfnamefont{M.}~\bibnamefont{Alvioli}}\ and\ \bibinfo
  {author} {\bibfnamefont{M.}~\bibnamefont{Strikman}},\ }%
  \bibfield{journal}{%
  \Doi{10.1103/PhysRevC.83.044905}{\bibinfo {journal} {Phys. Rev.}}\ }%
  \textbf{\bibinfo {volume} {C83}},\ \bibinfo {pages} {044905} (\bibinfo {year}
  {2011})%
  \bibAnnoteFile{NoStop}{Alvioli:2010yk}%
\bibitem{Teaney:2000cw}%
  \BibitemOpen
  \bibfield{author}{%
  \bibinfo {author} {\bibfnamefont{D.}~\bibnamefont{Teaney}}, \bibinfo {author}
  {\bibfnamefont{J.}~\bibnamefont{Lauret}},\ and\ \bibinfo {author}
  {\bibfnamefont{E.~V.}\ \bibnamefont{Shuryak}},\ }%
  \bibfield{journal}{%
  \bibinfo {journal} {Phys. Rev. Lett.}\ }%
  \textbf{\bibinfo {volume} {86}},\ \bibinfo {pages} {4783} (\bibinfo {year}
  {2001})%
  \bibAnnoteFile{NoStop}{Teaney:2000cw}%
\bibitem{Kolb:2003dz}%
  \BibitemOpen
  \bibfield{author}{%
  \bibinfo {author} {\bibfnamefont{P.~F.}\ \bibnamefont{Kolb}}\ and\ \bibinfo
  {author} {\bibfnamefont{U.~W.}\ \bibnamefont{Heinz}},\ }%
  in\ \emph{\bibinfo {booktitle} {Quark Gluon Plasma 3}},\ \bibinfo {editor}
  {edited by\ \bibinfo {editor} {\bibfnamefont{R.}~\bibnamefont{Hwa}}\ and\
  \bibinfo {editor} {\bibfnamefont{X.~N.}\ \bibnamefont{Wang}}}\ (\bibinfo
  {publisher} {World Scientific, Singapore},\ \bibinfo {year} {2004})\
  \Eprint{http://arxiv.org/abs/nucl-th/0305084}{arXiv:nucl-th/0305084}%
  \bibAnnoteFile{NoStop}{Kolb:2003dz}%
\bibitem{Hama:2005dz}%
  \BibitemOpen
  \bibfield{author}{%
  \bibinfo {author} {\bibfnamefont{Y.}~\bibnamefont{Hama}} \emph{et~al.},\ }%
  \bibfield{journal}{%
  \bibinfo {journal} {Nucl. Phys.}\ }%
  \textbf{\bibinfo {volume} {A774}},\ \bibinfo {pages} {169} (\bibinfo {year}
  {2006})%
  \bibAnnoteFile{NoStop}{Hama:2005dz}%
\bibitem{Huovinen:2006jp}%
  \BibitemOpen
  \bibfield{author}{%
  \bibinfo {author} {\bibfnamefont{P.}~\bibnamefont{Huovinen}}\ and\ \bibinfo
  {author} {\bibfnamefont{P.~V.}\ \bibnamefont{Ruuskanen}},\ }%
  \bibfield{journal}{%
  \Doi{10.1146/annurev.nucl.54.070103.181236}{\bibinfo {journal} {Ann. Rev.
  Nucl. Part. Sci.}}\ }%
  \textbf{\bibinfo {volume} {56}},\ \bibinfo {pages} {163} (\bibinfo {year}
  {2006})%
  \bibAnnoteFile{NoStop}{Huovinen:2006jp}%
\bibitem{Hirano:2005xf}%
  \BibitemOpen
  \bibfield{author}{%
  \bibinfo {author} {\bibfnamefont{T.}~\bibnamefont{Hirano}}, \bibinfo {author}
  {\bibfnamefont{U.~W.}\ \bibnamefont{Heinz}}, \bibinfo {author}
  {\bibfnamefont{D.}~\bibnamefont{Kharzeev}}, \bibinfo {author}
  {\bibfnamefont{R.}~\bibnamefont{Lacey}},\ and\ \bibinfo {author}
  {\bibfnamefont{Y.}~\bibnamefont{Nara}},\ }%
  \bibfield{journal}{%
  \Doi{10.1016/j.physletb.2006.03.060}{\bibinfo {journal} {Phys.Lett.}}\ }%
  \textbf{\bibinfo {volume} {B636}},\ \bibinfo {pages} {299} (\bibinfo {year}
  {2006})%
  \bibAnnoteFile{NoStop}{Hirano:2005xf}%
\bibitem{Hirano:2002ds}%
  \BibitemOpen
  \bibfield{author}{%
  \bibinfo {author} {\bibfnamefont{T.}~\bibnamefont{Hirano}}\ and\ \bibinfo
  {author} {\bibfnamefont{K.}~\bibnamefont{Tsuda}},\ }%
  \bibfield{journal}{%
  \bibinfo {journal} {Phys. Rev.}\ }%
  \textbf{\bibinfo {volume} {C66}},\ \bibinfo {pages} {054905} (\bibinfo {year}
  {2002})%
  \bibAnnoteFile{NoStop}{Hirano:2002ds}%
\bibitem{Bozek:2009ty}%
  \BibitemOpen
  \bibfield{author}{%
  \bibinfo {author} {\bibfnamefont{P.}~\bibnamefont{Bo\.zek}}\ and\ \bibinfo
  {author} {\bibfnamefont{I.}~\bibnamefont{Wyskiel}},\ }%
  \bibfield{journal}{%
  \Doi{10.1103/PhysRevC.79.044916}{\bibinfo {journal} {Phys. Rev.}}\ }%
  \textbf{\bibinfo {volume} {C79}},\ \bibinfo {pages} {044916} (\bibinfo {year}
  {2009})%
  \bibAnnoteFile{NoStop}{Bozek:2009ty}%
\bibitem{Ollitrault:1992bk}%
  \BibitemOpen
  \bibfield{author}{%
  \bibinfo {author} {\bibfnamefont{J.-Y.}\ \bibnamefont{Ollitrault}},\ }%
  \bibfield{journal}{%
  \Doi{10.1103/PhysRevD.46.229}{\bibinfo {journal} {Phys. Rev.}}\ }%
  \textbf{\bibinfo {volume} {D46}},\ \bibinfo {pages} {229} (\bibinfo {year}
  {1992})%
  \bibAnnoteFile{NoStop}{Ollitrault:1992bk}%
\bibitem{Iancu:2000hn}%
  \BibitemOpen
  \bibfield{author}{%
  \bibinfo {author} {\bibfnamefont{E.}~\bibnamefont{Iancu}}, \bibinfo {author}
  {\bibfnamefont{A.}~\bibnamefont{Leonidov}},\ and\ \bibinfo {author}
  {\bibfnamefont{L.~D.}\ \bibnamefont{McLerran}},\ }%
  \bibfield{journal}{%
  \Doi{10.1016/S0375-9474(01)00642-X}{\bibinfo {journal} {Nucl.Phys.}}\ }%
  \textbf{\bibinfo {volume} {A692}},\ \bibinfo {pages} {583} (\bibinfo {year}
  {2001})%
  \bibAnnoteFile{NoStop}{Iancu:2000hn}%
\bibitem{Ferreiro:2001qy}%
  \BibitemOpen
  \bibfield{author}{%
  \bibinfo {author} {\bibfnamefont{E.}~\bibnamefont{Ferreiro}}, \bibinfo
  {author} {\bibfnamefont{E.}~\bibnamefont{Iancu}}, \bibinfo {author}
  {\bibfnamefont{A.}~\bibnamefont{Leonidov}},\ and\ \bibinfo {author}
  {\bibfnamefont{L.}~\bibnamefont{McLerran}},\ }%
  \bibfield{journal}{%
  \Doi{10.1016/S0375-9474(01)01329-X}{\bibinfo {journal} {Nucl.Phys.}}\ }%
  \textbf{\bibinfo {volume} {A703}},\ \bibinfo {pages} {489} (\bibinfo {year}
  {2002})%
  \bibAnnoteFile{NoStop}{Ferreiro:2001qy}%
\bibitem{Drescher:2006pi}%
  \BibitemOpen
  \bibfield{author}{%
  \bibinfo {author} {\bibfnamefont{H.-J.}\ \bibnamefont{Drescher}}, \bibinfo
  {author} {\bibfnamefont{A.}~\bibnamefont{Dumitru}}, \bibinfo {author}
  {\bibfnamefont{A.}~\bibnamefont{Hayashigaki}},\ and\ \bibinfo {author}
  {\bibfnamefont{Y.}~\bibnamefont{Nara}},\ }%
  \bibfield{journal}{%
  \Doi{10.1103/PhysRevC.74.044905}{\bibinfo {journal} {Phys.Rev.}}\ }%
  \textbf{\bibinfo {volume} {C74}},\ \bibinfo {pages} {044905} (\bibinfo {year}
  {2006})%
  \bibAnnoteFile{NoStop}{Drescher:2006pi}%
\bibitem{Back:2001xy}%
  \BibitemOpen
  \bibfield{author}{%
  \bibinfo {author} {\bibfnamefont{B.~B.}\ \bibnamefont{Back}} \emph{et~al.}
  (\bibinfo {collaboration} {PHOBOS}),\ }%
  \bibfield{journal}{%
  \bibinfo {journal} {Phys. Rev.}\ }%
  \textbf{\bibinfo {volume} {C65}},\ \bibinfo {pages} {031901} (\bibinfo {year}
  {2002})%
  \bibAnnoteFile{NoStop}{Back:2001xy}%
\bibitem{Gyulassy:1996br}%
  \BibitemOpen
  \bibfield{author}{%
  \bibinfo {author} {\bibfnamefont{M.}~\bibnamefont{Gyulassy}}, \bibinfo
  {author} {\bibfnamefont{D.~H.}\ \bibnamefont{Rischke}},\ and\ \bibinfo
  {author} {\bibfnamefont{B.}~\bibnamefont{Zhang}},\ }%
  \bibfield{journal}{%
  \bibinfo {journal} {Nucl. Phys.}\ }%
  \textbf{\bibinfo {volume} {A613}},\ \bibinfo {pages} {397} (\bibinfo {year}
  {1997})%
  \bibAnnoteFile{NoStop}{Gyulassy:1996br}%
\bibitem{Takahashi:2009na}%
  \BibitemOpen
  \bibfield{author}{%
  \bibinfo {author} {\bibfnamefont{J.}~\bibnamefont{Takahashi}}, \bibinfo
  {author} {\bibfnamefont{B.}~\bibnamefont{Tavares}}, \bibinfo {author}
  {\bibfnamefont{W.}~\bibnamefont{Qian}}, \bibinfo {author}
  {\bibfnamefont{R.}~\bibnamefont{Andrade}}, \bibinfo {author}
  {\bibfnamefont{F.}~\bibnamefont{Grassi}}, \emph{et~al.},\ }%
  \bibfield{journal}{%
  \Doi{10.1103/PhysRevLett.103.242301}{\bibinfo {journal} {Phys.Rev.Lett.}}\ }%
  \textbf{\bibinfo {volume} {103}},\ \bibinfo {pages} {242301} (\bibinfo {year}
  {2009})%
  \bibAnnoteFile{NoStop}{Takahashi:2009na}%
\bibitem{Sorensen:2011xw}%
  \BibitemOpen
  \bibfield{author}{%
  \bibinfo {author} {\bibfnamefont{P.}~\bibnamefont{Sorensen}},\ }%
  \bibfield{journal}{%
  \Doi{10.1016/j.nuclphysa.2011.02.046}{\bibinfo {journal} {Nucl.Phys.}}\ }%
  \textbf{\bibinfo {volume} {A855}},\ \bibinfo {pages} {229} (\bibinfo {year}
  {2011})%
  \bibAnnoteFile{NoStop}{Sorensen:2011xw}%
\bibitem{Aggarwal:2001aa}%
  \BibitemOpen
  \bibfield{author}{%
  \bibinfo {author} {\bibfnamefont{M.~M.}\ \bibnamefont{Aggarwal}}
  \emph{et~al.} (\bibinfo {collaboration} {WA98}),\ }%
  \bibfield{journal}{%
  \Doi{10.1103/PhysRevC.65.054912}{\bibinfo {journal} {Phys. Rev.}}\ }%
  \textbf{\bibinfo {volume} {C65}},\ \bibinfo {pages} {054912} (\bibinfo {year}
  {2002})%
  \bibAnnoteFile{NoStop}{Aggarwal:2001aa}%
\bibitem{Adare:2008ns}%
  \BibitemOpen
  \bibfield{author}{%
  \bibinfo {author} {\bibfnamefont{A.}~\bibnamefont{Adare}} \emph{et~al.}
  (\bibinfo {collaboration} {PHENIX}),\ }%
  \bibfield{journal}{%
  \Doi{10.1103/PhysRevC.78.044902}{\bibinfo {journal} {Phys. Rev.}}\ }%
  \textbf{\bibinfo {volume} {C78}},\ \bibinfo {pages} {044902} (\bibinfo {year}
  {2008})%
  \bibAnnoteFile{NoStop}{Adare:2008ns}%
\bibitem{Gazdzicki:2008kk}%
  \BibitemOpen
  \bibfield{author}{%
  \bibinfo {author} {\bibfnamefont{M.}~\bibnamefont{Gazdzicki}} (\bibinfo
  {collaboration} {NA61/SHINE Collaboration}),\ }%
  \bibfield{journal}{%
  \Doi{10.1088/0954-3899/36/6/064039}{\bibinfo {journal} {J.Phys.G}}\ }%
  \textbf{\bibinfo {volume} {G36}},\ \bibinfo {pages} {064039} (\bibinfo {year}
  {2009})%
  \bibAnnoteFile{NoStop}{Gazdzicki:2008kk}%
\bibitem{:2009vy}%
  \BibitemOpen
  \bibfield{author}{%
  \bibinfo {author} {\bibfnamefont{T.}~\bibnamefont{Anticic}} \emph{et~al.}
  (\bibinfo {collaboration} {NA49 Collaboration, NA61/SHINE Collaboration}),\
  }%
  \bibfield{journal}{%
  \bibinfo {journal} {PoS}\ }%
  \textbf{\bibinfo {volume} {EPS-HEP2009}},\ \bibinfo {pages} {030} (\bibinfo
  {year} {2009})%
  \bibAnnoteFile{NoStop}{:2009vy}%
\bibitem{:2009vj}%
  \BibitemOpen
  \bibfield{author}{%
  \bibinfo {author} {\bibfnamefont{N.}~\bibnamefont{Abgrall}} \emph{et~al.}
  (\bibinfo {collaboration} {NA61 Collaboration}),\ }%
  \bibfield{journal}{%
  \bibinfo {journal} {PoS}\ }%
  \textbf{\bibinfo {volume} {CPOD2009}},\ \bibinfo {pages} {049} (\bibinfo
  {year} {2009})%
  \bibAnnoteFile{NoStop}{:2009vj}%
\bibitem{Gazdzicki:2011fx}%
  \BibitemOpen
  \bibfield{author}{%
  \bibinfo {author} {\bibfnamefont{M.}~\bibnamefont{Gazdzicki}} (\bibinfo
  {collaboration} {NA49 Collaboration, NA61/SHINE Collaboration})}%
   (\bibinfo {year} {2011}),\ \bibinfo {note} {* Temporary entry *},\
  \Eprint{http://arxiv.org/abs/1107.2345}{arXiv:1107.2345 [nucl-ex]}%
  \bibAnnoteFile{NoStop}{Gazdzicki:2011fx}%
\bibitem{Gazdzicki:1322135}%
  \BibitemOpen
  \bibfield{author}{%
  \bibinfo {author} {\bibfnamefont{M.}~\bibnamefont{Gazdzicki}},\ }%
  \emph{\bibinfo {title} {The 2010 test of secondary light ion beams}},\
  \bibinfo {type} {Tech. Rep.}\ \bibinfo {number} {CERN-SPSC-2011-005.
  SPSC-SR-077}\ (\bibinfo {institution} {CERN},\ \bibinfo {address} {Geneva},\
  \bibinfo {year} {2011})%
  \bibAnnoteFile{NoStop}{Gazdzicki:1322135}%
\bibitem{:2009dqa}%
  \BibitemOpen
  \bibfield{author}{%
  \bibinfo {author} {\bibfnamefont{B.~I.}\ \bibnamefont{Abelev}} \emph{et~al.}
  (\bibinfo {collaboration} {STAR}),\ }%
  \bibfield{journal}{%
  \Doi{10.1103/PhysRevLett.103.172301}{\bibinfo {journal} {Phys. Rev. Lett.}}\
  }%
  \textbf{\bibinfo {volume} {103}},\ \bibinfo {pages} {172301} (\bibinfo {year}
  {2009})%
  \bibAnnoteFile{NoStop}{:2009dqa}%
\bibitem{Back:2006id}%
  \BibitemOpen
  \bibfield{author}{%
  \bibinfo {author} {\bibfnamefont{B.~B.}\ \bibnamefont{Back}} \emph{et~al.}
  (\bibinfo {collaboration} {PHOBOS}),\ }%
  \bibfield{journal}{%
  \Doi{10.1103/PhysRevC.74.011901}{\bibinfo {journal} {Phys. Rev.}}\ }%
  \textbf{\bibinfo {volume} {C74}},\ \bibinfo {pages} {011901} (\bibinfo {year}
  {2006})%
  \bibAnnoteFile{NoStop}{Back:2006id}%
\bibitem{Bialas:2011bz}%
  \BibitemOpen
  \bibfield{author}{%
  \bibinfo {author} {\bibfnamefont{A.}~\bibnamefont{Bialas}}, \bibinfo {author}
  {\bibfnamefont{A.}~\bibnamefont{Bzdak}},\ and\ \bibinfo {author}
  {\bibfnamefont{K.}~\bibnamefont{Zalewski}}}%
   (\bibinfo {year} {2011}),\
  \Eprint{http://arxiv.org/abs/1107.1215}{arXiv:1107.1215 [hep-ph]}%
  \bibAnnoteFile{NoStop}{Bialas:2011bz}%
\end{thebibliography}

\end{document}